\documentclass[lettersize,journal]{IEEEtran}

\usepackage{stackengine}
\usepackage{amsmath,amssymb,amsfonts}
\usepackage{algorithmic}
\usepackage[ruled,vlined]{algorithm2e}
\usepackage{etoolbox} 
\usepackage[dvipsnames]{xcolor}
\usepackage{colortbl}
\usepackage{pgfplots}
\usepackage{verbatim}
\pgfplotsset{width=7cm,compat=1.8}
\usepgfplotslibrary{fillbetween}
\usepackage{xcolor}
\usepackage{tikz}
\usepackage{tikzscale}
\usetikzlibrary{patterns}
\pgfplotsset{every tick/.style={black,}}
\usepackage{subcaption}
\usepackage{caption}
\usepackage{amsmath}
\usepackage{multirow}
\usepackage{multicol}
\usepackage{url}
\usepackage{pgfplotstable}
\usepackage{array}
\usepgfplotslibrary{groupplots}
\usetikzlibrary{patterns}
\usepackage[normalem]{ulem}

\usepackage[super]{nth}
\usepackage{tabularx}
\usepackage{bbding}
\usepackage{booktabs}
\usepackage{adjustbox}
\usepackage{graphicx}
\usepackage{authblk}

\patchcmd{\algocf@makecaption@ruled}{\hsize}{\textwidth}{}{} 
\patchcmd{\@algocf@start}{-1.5em}{0em}{}{} 

\pgfplotsset{compat=1.11,
        /pgfplots/ybar legend/.style={
        /pgfplots/legend image code/.code={%
        \draw[##1,/tikz/.cd,bar width=8pt,yshift=-0.2em,bar shift=0pt]
                plot coordinates {(0cm,0.8em)};},
},
}

\newcommand\xrowht[2][0]{\addstackgap[.5\dimexpr#2\relax]{\vphantom{#1}}}

\newcolumntype{b}{>{\columncolor{blue!10}}c}
\newcolumntype{y}{>{\columncolor{yellow!10}}c}
\newcolumntype{d}{>{\columncolor{red!7}}c}

\newcolumntype{C}{>{\centering\arraybackslash}X}
\newcolumntype{D}{>{\hsize=\dimexpr2\hsize+2\tabcolsep+\arrayrulewidth\relax}C}

\usetikzlibrary{plotmarks} 

\newcommand{\pltheight}{5.5cm}
\newcommand{\norm}[1]{\left\lVert#1\right\rVert}

\begin{document}

\title{VERITAS: \uline{V}erifying the P\uline{er}formance of A\uline{I}-native \uline{T}ransceiver \uline{A}ctions in Base-\uline{S}tations}

\author[*]{Nasim Soltani}
\author[$\dag$]{Michael L{\"o}hning}
\author[*]{Kaushik Chowdhury}
\affil[*]{Electrical and Computer Engineering Department, The University of Texas at Austin, Austin, TX, USA}
\affil[$\dag$]{National Instruments - Test and Measurement Group of Emerson, Dresden, Germany}
\affil[ ]{nasim.soltani@utexas.edu, michael.loehning@emerson.com, kaushik@utexas.edu}

\markboth{This work is accepted at IEEE Transactions on Mobile Computing, November 2025.}%
{}

\maketitle

\begin{abstract}
Artificial Intelligence (AI)-native receivers provide lower bit error rate (BER) compared to the traditional receiver, if they are deployed on the same data distribution as their training set. A major research problem is the uncertainty of whether a particularly trained AI-native receiver maintains its superior performance over the traditional receiver in different deployment environments. To this end, we propose VERITAS as a joint measurement-recovery post deployment framework for AI-native transceivers that continuously looks for distribution shifts in the received pilots and triggers finite re-training spurts. VERITAS leverages a novel out-of-distribution algorithm to detect potential changes in the channel profile, transmitter speed, and delay spread. As soon as such a change is detected, a traditional (reference) receiver is activated, which runs for a period of time in parallel to the AI-native receiver. Finally, VERTIAS compares the bit probabilities of the AI-native and the reference receivers for the same received data inputs, and decides whether or not a retraining process needs to be initiated. Our evaluations reveal that VERITAS can detect changes in the channel profile, transmitter speed, and delay spread with 99\%, 97\%, and 78\% accuracies, respectively, followed by timely initiation of retraining for 86\%, 93.3\%, and 94.8\% of inputs in channel profile, transmitter speed, and delay spread test sets, respectively.  

\end{abstract}
 

\vspace{-2mm}
\section{Introduction}\label{sec:intro}
Artificial intelligence native air interface (AI-AI) offers a fully AI-based interface for next-generation wireless communications, where AI is integrated in both data and control paths~\cite{ai-ai,ericsson-ai-ai}. AI-AI provides a myriad of flexibilities and opportunities for physical layer design, including but not limited to: merging data decoding and application in the physical layer, providing flexibility in the choice of waveform with respect to the radio hardware and environment constraints, obviating costly hardware implementation for each individual processing block by being fully AI-based, reduction in standardization need, and the possibility of physical and media access control (MAC) layer fusion~\cite{ai-ai,ericsson-ai-ai}. Furthermore, as data decoding and interpretation happens through neural networks (NNs) that have learned to map received data to originally transmitted bits, AI-based receivers previously showed to yield lower bit error rate (BER) compared to receivers with traditional signal processing blocks. Examples of such demonstration are shown with over-the-air data in~\cite{nasim-spinn,bahar-spinn} and on real hardware in~\cite{ni-white-deeprx}.

\begin{figure}[t!!!]
    \centering
    \includegraphics[width=0.9
    \linewidth]{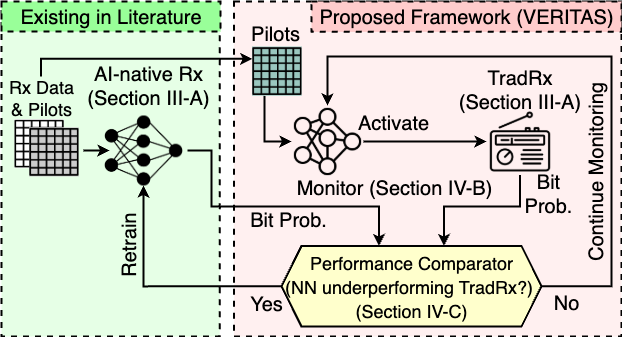}
    \caption{VERITAS for verifying the performance of AI-native Receiver. The Monitor continuously scans the wireless channel, and as soon as it detects a change, it activates TradRx. Bit probabilities of AI-native receiver and TradRx are compared to find the underperforming receiver. If necessary, the AI-native receiver is updated through retraining to adapt to the new wireless channel.}
    \label{fig:overview}
\end{figure}

\noindent \textbf{Problem.} Despite the several benefits that AI-AI provides for 6G communications, this newly proposed paradigm faces key challenges that need to be addressed before its successful deployment in 6G systems. For example, since wireless channel is a major contributor in the NN-based receiver performance, \textit{verifying and maintaining the performance} of the AI-native receiver becomes a real challenge. NNs might not deliver the expected performance if they are deployed under wireless channels different from what they were trained in, as evident in several different examples: Recent work~\cite{al2020exposing,nasim-augmentation} show that NN-based radio frequency (RF) fingerprinting accuracy drops drastically when training channels are different from deployment channels. Authors in~\cite{nasim-spinn} show retraining is necessary to maintain the performance of an NN-based channel estimator in new environments. Authors in~\cite{modulation-classification} show that automatic modulation classification performance drops when the wireless channel changes, and propose transfer learning as an effective light retraining technique. The critical role of the wireless channel in AI-based wireless systems renders the performance verification and maintenance of AI-AI necessary.

\noindent \textbf{Preliminaries.} As shown on the left side of Fig.~\ref{fig:overview}, we assume an NN-based receiver (a.k.a., AI-native receiver) that is responsible for converting received wireless signals (i.e., Rx Data \& Pilots) to bits. Such models are well-explored in the literature in a number of prior works~\cite{experimental,Pilotless,CP-free,deep-waveform,data-driven-letter,gao2018comnet,nasim-spinn,bahar-spinn}. Due to practical limitations, the training set of the NN-based receiver cannot contain all possible data variations or signals recorded under all possible channels encountered in the real world. Instead, the NN-based receiver is trained on a number of channel profiles and mobility conditions. While it performs well under the seen configurations~\cite{deeprx,ni-white-deeprx}, it is an open question whether it may suffer from a performance drop if deployed under a new wireless channel.

\noindent \textbf{Limitation of Existing Solutions.} To ensure maintained performance of the NN-based receiver, authors in~\cite{adaptive-NN-receiver} propose periodic retraining as a measure to adapt the receiver to new channel conditions. However, fixed-time-interval retraining imposes significant training computational complexity and requires dedicated computation resources. Furthermore, retraining an NN-based receiver in the field requires collecting signals that are \emph{labeled} with transmitter-side bits under the new channel, which is not a trivial task. Therefore, unnecessary retraining must be avoided. Retraining must be applied only when we know that the performance of the NN-based receiver has definitely dropped compared to the traditional receiver, under the new channel conditions.

\noindent \textbf{Proposed Solution.} On the right side of Fig.~\ref{fig:overview}, we propose VERITAS as a framework for verifying the performance of AI-native receiver in the field and limit its retraining to only necessary situations. VERITAS has 3 components: the Monitor, the traditional receiver (TradRx), and the Performance Comparator. The Monitor runs continuously in parallel to the AI-native receiver to observe the wireless channel and detect potential changes in the channel profile type, transmitter speed, or delay spread. As soon as such a change is detected, TradRx is activated and used as a comparison point against the AI-native receiver. The Performance Comparator compares output bit probabilities of the two receivers and determines the underperforming receiver. Notably, this step does not require the true bit labels or BER calculations. If the AI-native receiver is identified as the underperforming receiver, a retraining process is initiated to lightly retrain the AI-native receiver to ensure its maintained superior performance over the traditional receiver. The proposed Performance Comparator in VERITAS is able to operate on encoded as well as raw (i.e., uncoded) bits, which obviates the need for a costly decoding block within the proposed framework.

\noindent \textbf{Contributions.} Our contributions are as follows:
\begin{itemize}
    \item We propose VERITAS as a framework for verifying the performance of an AI-native receiver, to ensure its maintained superior decoding performance compared to traditional receiver (Section~\ref{sec:method-system}).
    \item To demonstrate VERITAS works for generic AI-native receivers, we choose a widely used NN-based 5G receiver called DeepRx~\cite{deeprx} as our AI-native receiver (Section~\ref{sec:prelim-pipelines}), which is designed to give lower BER compared to TradRx. We extensively analyze DeepRx performance for different training and test set configurations, and determine configurations where DeepRx yields higher BER compared to TradRx (Section~\ref{sec:prelim-drop}).
    \item We propose a wireless channel change detector called Monitor designed as a custom NN cascaded with a novel out-of-distribution (OOD) algorithm based on K-nearest neighbor (KNN)~\cite{knn}. The Monitor processes received 5G pilots and identifies any potential changes in the channel profile, transmitter speed, and delay spread. (Section~\ref{sec:method-monitor}). 
    \item We propose an analytical method based on histogram binning to compare the output bit probabilities of the AI-native receiver against those of TradRx as reference. The proposed Performance Comparator compares output bit probabilities at the deployment phase and without having the true bit labels. This comparison determines if the AI-native receiver is underperforming with respect to the reference, which initiates a retraining process (Section~\ref{sec:method-comparator}). 
    \item We pledge to publicly release our code~\cite{veritas-repo} for VERITAS including pipelines for the Monitor and the Performance Comparator, upon the acceptance of this paper. 
\end{itemize}
\section{Related Work}\label{sec:related}
In this section, we summarize the closest related work in three different areas of AI-native receiver performance maintenance (Section~\ref{sec:related-maintenace}), wireless channel change detection (Section~\ref{sec:related-environment}), and OOD detection (Section~\ref{sec:related-ood}).

\vspace{-3mm}
\subsection{AI-native Receiver Performance Maintenance}\label{sec:related-maintenace}
The issue of performance drop in the NN-based receivers due to channel variations has been studied extensively. Authors in~\cite{adaptive-NN-receiver} propose a fixed time interval (periodic) retraining technique to adapt NN-based orthogonal frequency division multiplexing (OFDM) receivers to occasionally changing channel conditions. Naive periodic retraining is a way of maintaining performance of an NN-based receiver, however, it periodically imposes often unnecessary training computational complexity to the system as well as wastage of data frames that are used as the retraining dataset. 
Authors in~\cite{mauro-channelEst} propose a de-noising approach during training for learning OFDM channel coefficients. They construct their training set out of estimated channel coefficients of low noise signals, but dynamically add additive white Gaussian noise (AWGN) to inputs during training. This method makes the NN-based channel estimator robust to changes in the noise level, however, this does not solve the problem of transitioning between different wireless channels between training and deployment phases.

\vspace{-3mm}
\subsection{Wireless Channel Change Detection}\label{sec:related-environment}
Authors in~\cite{environment-classifier} use the channel state information (CSI) of IEEE 802.11p signals for environment identification in V2V communication. They consider 5 different environments of rural line-of-sight (LOS), urban LOS, urban non-line-of-sight (NLOS), highway LOS, and highway NLOS for V2V communication and perform a multi-class classification using a deep convolutional NN, KNN, support vector machine, random forest, and Gaussian naive Bayes algorithms. They show superior performance of the NN over the other algorithms, however, they do not go beyond the fixed training set environments and do not show any method for identifying new environments. Authors in~\cite{speed-classifier} classify user speeds in a 5G network using the reference signal received power (RSRP) passed through a deep NN. They categorize speeds between 0 and $\infty$ km/h into 8 non-overlapping classes, with the last class spanning from 90 km/h to $\infty$. This categorization encompasses all the possible speeds, however, known-class speed classification without OOD detection does not satisfy the requirements in our proposed AI-native receiver maintenance framework.

\vspace{-3mm}

\subsection{OOD Detection}\label{sec:related-ood}
Detecting OOD samples is a well-investigated problem in machine learning~\cite{open-ood,ood-survey}. In wireless communications, autoencoders have been vastly used for OOD detection. In such methods an autoencoder is trained to reconstruct an input, and the reconstruction error for known in-distribution (ID) inputs are averaged and recorded as a reference. At the deployment phase, all unknown inputs are fed to the autoencoder and their reconstruction errors are compared against the reference error. If the reconstruction error of the unknown test input is larger than the reference, the input is identified as an OOD input. Authors in~\cite{ood-wireless} use a variational autoencoder and study the signal reconstruction error for identifying OOD modulation schemes.
The disadvantage of autoencoder-based OOD detection is that completely different pipelines are needed for OOD detection and ID data classification. On the other hand, classification-based OOD detection methods provide a unified pipeline for both tasks. Authors in~\cite{robinson2021novel} detect unseen devices in the well-known RFMLS~\cite{rfmls} WiFi and ADS-B datasets using a classification-based OOD detection method. They train their classifier NN with a custom loss function that has 3 components of intra-centroid loss, nearest neighbor loss and a final loss component that pushes the cluster centers away to spread in the space. However, their proposed method requires exposure of the NN to out-of-library devices (classes) that are not categorized into meaningful classes during training. Authors in~\cite{RF-fingerprinting-lora} include a feature-based new device detection in their proposed LoRa RF fingerprinting scheme. They calculate the average of distances of test features from all of its $K$ nearest neighbors, and compare it to a predefined $\lambda$ value, and decide if the device is OOD or has been seen during training. This averaging process causes information loss and might degrade OOD detection performance. Furthermore, relying exclusively on distances from neighbors limits the methods to ID clusters that are dense in the center and scattered around the edges.

In the rest of this paper, we introduce a widely used NN-based receiver as our example AI-native receiver, and explore its performance for different training and test set configurations. Then, we describe and evaluate VERITAS as a framework for verifying the performance of this AI-native receiver to avoid its periodic and often unnecessary retraining. 
\section{Preliminaries}\label{sec:preliminaries}
In this section, we briefly describe different pipelines for data generation, the traditional receiver, and the NN-based receiver that we use in this paper as our example AI-native receiver (Section~\ref{sec:prelim-pipelines}). We explore the performance of the NN-based receiver in different training and test set configurations, and attempt to find cases where DeepRx shows a higher BER compared to TradRx and refer to them as \emph{performance drop cases} (Section~\ref{sec:prelim-drop}).

\vspace{-3mm}
\subsection{Data Generation, TradRx, and AI-native Receiver Pipelines}\label{sec:prelim-pipelines}
\noindent \textbf{Data Generation Pipeline.} We generate data using \texttt{Sionna} libraries by synthesizing transmitter 5G radio frames containing random bits and passing them through 3GPP 38.901 tap-delay line channel models tdl\_a, tdl\_b, tdl\_c, tdl\_d, and tdl\_e, that are implemented and available within \texttt{Sionna}. After the simulated channel, we also use the \texttt{Sionna} API \texttt{AWGN()} to add specific levels of noise to the data.

\noindent \textbf{TradRx.} Our traditional receiver that we refer to as TradRx, is based on least square (LS) channel estimation and linear minimum mean square error (LMMSE) equalization. To implement TradRx, different \texttt{Sionna} classes and functions including \texttt{OFDMDemodulator}, \texttt{LSChannelEstimator}, \texttt{LMMSEEqualizer}, and \texttt{Demapper} are used. The implemented TradRx is used as a reference for benchmarking the performance of AI-native receiver for different dataset configurations. 

\noindent \textbf{The 5G AI-native Receiver: DeepRx.} As our AI-native receiver, we adopt a widely used fully convolutional 5G receiver with 672k parameters called DeepRx~\cite{deeprx}. DeepRx interprets frequency domain I/Q samples in 5G subframes to their corresponding softbits (a.k.a., log likelyhood ratios (LLRs)). The input of DeepRx is a (14, 72, 6) tensor that contains real and imaginary parts of the frequency domain received 5G subframes, raw estimated channel coefficients, and transmitter-side pilot symbols, stacked together in the last dimension. More details about DeepRx architecture can be found in~\cite{deeprx}.
We note that as the error correction block is not part of the DeepRx NN in~\cite{deeprx}, we do not include this block in the implementation of either DeepRx or TradRx. Therefore, without losing generality of our proposed method, all the BER results reported in the rest of this paper are reported for \emph{uncoded} bits.

\vspace{-2mm}
\subsection{Exploration of DeepRx Performance Compared to TradRx}\label{sec:prelim-drop}
\begin{table*}[t!!!]
    \centering
    \begin{tabular}{|l|l|l|l|c|c|c|}
        \hline
        \xrowht[()]{7pt}
        \# & Experiment name & BER vs. Eb/N0 & Dataset & Channel Profile & Transmitter Speed (m/s) & Delay Spread (ns) \\
        \hline \hline
        \xrowht[()]{7pt}
        \multirow{2}{*}{1} & \multirow{2}{*}{Channel Profile - Exp. 1} & \multirow{2}{*}{Fig.~\ref{fig:prelim-channel-pass}} & Training & \cellcolor{ForestGreen!10} tdl\_a & 10 & 400 \\
        & & & Test & \cellcolor{ForestGreen!10} tdl\_a, tdl\_b, tdl\_c, tdl\_d, tdl\_e & 10 & 400 \\
        \hline
        \xrowht[()]{7pt}
        \multirow{2}{*}{2} & \multirow{2}{*}{Channel Profile - Exp. 2} & \multirow{2}{*}{Fig.~\ref{fig:prelim-channel-drop}} & Training & \cellcolor{ForestGreen!10}tdl\_d & 10 & 400 \\
        & & & Test & \cellcolor{ForestGreen!10} tdl\_a, tdl\_b, tdl\_c, tdl\_d, tdl\_e & 10 & 400 \\
        \hline
        \xrowht[()]{7pt}
        \multirow{2}{*}{3} & \multirow{2}{*}{Transmitter Speed - Exp. 1} & \multirow{2}{*}{Fig.~\ref{fig:prelim-speed-trained-high}} & Training & tdl\_d & \cellcolor{red!10} 18, 19, 20 & 400 \\
        & & & Test & tdl\_d & \cellcolor{red!10} 1, 16, 17 & 400 \\
        \hline
        \xrowht[()]{7pt}
        \multirow{2}{*}{4} & \multirow{2}{*}{Transmitter Speed - Exp. 2} & \multirow{2}{*}{Fig.~\ref{fig:prelim-speed-trained-low}} & Training & tdl\_d & \cellcolor{red!10} 0, 1, 2 & 400 \\
        & & & Test & tdl\_d & \cellcolor{red!10} 3, 4, 20 & 400 \\
        \hline
        \xrowht[()]{7pt}
        \multirow{2}{*}{5} & \multirow{2}{*}{Delay Spread - Exp. 1} & \multirow{2}{*}{Fig.~\ref{fig:prelim-delay-trained-high}} & Training & tdl\_b & 2 & \cellcolor{blue!10} 400, 450, 500 \\
        & & & Test & tdl\_b & 2 & \cellcolor{blue!10} 10, 50, 80 \\
        \hline
        \xrowht[()]{7pt}
        \multirow{2}{*}{6} & \multirow{2}{*}{Delay Spread - Exp. 2} & \multirow{2}{*}{Fig.~\ref{fig:prelim-delay-trained-low}} & Training & tdl\_b & 2 & \cellcolor{blue!10} 10, 50, 80 \\
        & & & Test & tdl\_b & 2 & \cellcolor{blue!10} 100, 200, 400 \\
        \hline
    \end{tabular}
    \caption{Summary of preliminary experiments description. Colored cells show the parameters that are variant between the training and test set datasets for each experiment.}
    \label{tab:exp_description}
\end{table*}

\begin{figure*}
\captionsetup{width=0.45\linewidth}
\centering
    \begin{minipage}{0.49\textwidth}
\centering
\begin{tikzpicture}
        \begin{axis}[
            xlabel = Eb/N0 (dB),
            ylabel = BER (log),
            title={Channel Profile - Exp. 1},
            ymin=0.000025,
            ymax=0.24,
            ymode = log,
            xmin=0,
            xmax=20,
            height=\pltheight,
            width=\linewidth,
            legend style={at={(0,0)},
            fill=gray!5,
            fill opacity=0.6,
            text opacity=1,
            draw=none,
            font=\small,
            legend cell align=left,
                anchor=south west,legend columns=2},
            ymajorgrids=true,
            ytick = {0.1,0.01,0.001,0.0001},
            yticklabels = {$10^{-1}$,$10^{-2}$,$10^{-3}$,$10^{-4}$},
            grid style={line width=1pt,draw=gray!50},
        ]
        \addplot [thick,blue,mark=|,mark size=4pt]
        coordinates{ 
        (0,0.1843685061)(2,0.1450125575)(4,0.1053501144)(6,0.07742738724)(8,0.05286916718)(10,0.03318211064)(12,0.02097505517)(14,0.01415238902)(16,0.008312722668)(18,0.005879166536)(20,0.003674000036)
        (0,0.1932290494441986)(2,0.15367645025253296)(4,0.11686116456985474)(6,0.08601866662502289)(8,0.0584971122443676)(10,0.03957561030983925)(12,0.027062777429819107)(14,0.01807405613362789)(16,0.011978666298091412)(18,0.007968555204570293)(20,0.00527055561542511)
        };
        \addlegendentry{DeepRx, tdl\_a}
        \addplot [thick,dashed,blue,mark=|,mark options=solid,mark size=4pt]
        coordinates{ 
        (0,0.2272983333333335)(2,0.18468449999999992)(4,0.14344011111111116)(6,0.10800183333333327)(8,0.07497283333333327)(10,0.05182811111111114)(12,0.03554816666666664)(14,0.02369677777777776)(16,0.015545777777777787)(18,0.010231222222222225)(20,0.006585166666666667)
        };
        \addlegendentry{TradRx, tdl\_a}
        \addplot [thick,red,mark=o]
        coordinates{
        (0,0.19926400482654572)(2,0.15353511273860931)(4,0.11883799731731415)(6,0.08716761320829391)(8,0.06201572343707085)(10,0.04358605667948723)(12,0.028697500005364418)(14,0.01928822137415409)(16,0.012307999655604362)(18,0.008632278069853783)(20,0.0053858887404203415)
        };
        \addlegendentry{DeepRx, tdl\_b}
        \addplot [thick,red,dashed,mark=o,mark options=solid]
        coordinates{
        (0,0.23342200000000016)(2,0.18431683333333335)(4,0.1455022222222222)(6,0.10905427777777775)(8,0.07914305555555551)(10,0.056140166666666685)(12,0.037428833333333356)(14,0.025219555555555582)(16,0.015948166666666673)(18,0.011043333333333327)(20,0.006748666666666673)
        };
        \addlegendentry{TradRx, tdl\_b}
        \addplot [thick,orange,mark=x,mark size=4pt]
        coordinates{
        (0,0.19582949578762054)(2,0.1546970009803772)(4,0.11945144087076187)(6,0.08682116866111755)(8,0.06328783184289932)(10,0.04194272309541702)(12,0.02792000025510788)(14,0.018251944333314896)(16,0.012187832966446877)(18,0.008063111454248428)(20,0.005580611061304808)
        };
        \addlegendentry{DeepRx, tdl\_c}
        \addplot [thick,orange,mark=x,dashed,mark options=solid,mark size=4pt]
        coordinates{
        (0,0.22911977777777773)(2,0.1848350555555556)(4,0.14547294444444436)(6,0.10803772222222234)(8,0.0798918333333334)(10,0.05392533333333337)(12,0.03622377777777781)(14,0.023701222222222217)(16,0.015734722222222212)(18,0.010275722222222231)(20,0.006853444444444443)
        };
        \addlegendentry{TradRx, tdl\_c}
        \addplot [thick,Plum,mark=*]
        coordinates{
        (0,0.15738427639007568)(2,0.11478455364704132)(4,0.07631133496761322)(6,0.04370177909731865)(8,0.02109716646373272)(10,0.008474444039165974)(12,0.0031547776889055967)(14,0.0010432222625240684)(16,0.00033544443431310356)(18,0.00016349999350495636)(20,6.088888767408207e-05)
        };
        \addlegendentry{DeepRx, tdl\_d}
        \addplot [thick,Plum,mark=*,dashed,mark options=solid]
        coordinates{
        (0,0.1895611666666668)(2,0.1426593333333332)(4,0.09959444444444447)(6,0.06146655555555555)(8,0.0327201111111111)(10,0.014666111111111114)(12,0.005818611111111107)(14,0.001974555555555554)(16,0.0005868888888888885)(18,0.00023972222222222282)(20,7.52222222222223e-05)
        };
        \addlegendentry{TradRx, tdl\_d}
        \addplot [thick,ForestGreen,mark=square*]
        coordinates{
        (0,0.1560332179069519)(2,0.11274150013923645)(4,0.07611633092164993)(6,0.04299205541610718)(8,0.02098538912832737)(10,0.008672777563333511)(12,0.0029841666109859943)(14,0.0008989444468170404)(16,0.0003374444495420903)(18,6.394444790203124e-05)(20,3.4111111745005473e-05)
        };
        \addlegendentry{DeepRx, tdl\_e}
        \addplot [thick,ForestGreen,mark=square*,dashed,mark options=solid]
        coordinates{
        (0,0.1892956666666667)(2,0.14183172222222226)(4,0.10085349999999996)(6,0.062195944444444436)(8,0.0336687222222222)(10,0.015562722222222226)(12,0.006011111111111114)(14,0.0019797777777777787)(16,0.0006874444444444443)(18,0.00014388888888888923)(20,6.400000000000008e-05)
        };
        \addlegendentry{TradRx, tdl\_e}
        \end{axis}
    \end{tikzpicture}
    \caption{BER vs. Eb/N0 in Channel Profile - Exp. 1, where DeepRx is trained on tdl\_a (NLOS) channel. During test, DeepRx outperforms TradRx in LOS and NLOS channel profiles, and therefore, no \emph{performance drop case} is observed.}
    \label{fig:prelim-channel-pass}
    \end{minipage}
    \begin{minipage}{0.49\textwidth}
\centering
\begin{tikzpicture}
        \begin{axis}[
            xlabel = Eb/N0 (dB),
            ylabel = BER (log),
            title={Channel Profile - Exp. 2},
            ymin=0.00003,
            ymax=0.24,
            ymode = log,
            xmin=0,
            xmax=20,
            height=\pltheight,
            width=\linewidth,
            legend style={at={(0,0)},
            fill=gray!5,
            fill opacity=0.6,
            text opacity=1,
            draw=none,
            font=\small,
            legend cell align=left,
                anchor=south west,legend columns=2},
            ymajorgrids=true,
            ytick = {0.1,0.01,0.001,0.0001},
            yticklabels = {$10^{-1}$,$10^{-2}$,$10^{-3}$,$10^{-4}$},
            grid style={line width=1pt,draw=gray!50},
        ]
        \addplot [thick,blue,mark=|,mark size=4pt]
        coordinates{ 
        (0,0.202290386)(2,0.1630704999)(4,0.126650393)(6,0.09576461464)(8,0.06745450199)(10,0.04869350046)(12,0.03561788797)(14,0.02638277784)(16,0.02087666653)(18,0.01662727818)(20,0.01394133363)
        };
        \addlegendentry{DeepRx, tdl\_a}
        \addplot [thick,dashed,blue,mark=|,mark options=solid,mark size=4pt]
        coordinates{ 
        (0,0.2272983333)(2,0.1846845)(4,0.1434401111)(6,0.1080018333)(8,0.07497283333)(10,0.05182811111)(12,0.03554816667)(14,0.02369677778)(16,0.01554577778)(18,0.01023122222)(20,0.006585166667)
        };
        \addlegendentry{TradRx, tdl\_a}
        \addplot [thick,red,mark=o]
        coordinates{
        (0,0.2074578851)(2,0.1625667214)(4,0.1281982213)(6,0.09643471986)(8,0.07092277706)(10,0.05227416754)(12,0.03735638782)(14,0.02769749984)(16,0.02030066587)(18,0.01714911126)(20,0.01390533336)
        };
        \addlegendentry{DeepRx, tdl\_b}
        \addplot [thick,red,dashed,mark=o,mark options=solid]
        coordinates{
        (0,0.233422)(2,0.1843168333)(4,0.1455022222)(6,0.1090542778)(8,0.07914305556)(10,0.05614016667)(12,0.03742883333)(14,0.02521955556)(16,0.01594816667)(18,0.01104333333)(20,0.006748666667)
        };
        \addlegendentry{TradRx, tdl\_b}
        \addplot [thick,orange,mark=x,mark size=4pt]
        coordinates{
        (0,0.2028820515)(2,0.1622428298)(4,0.1272504926)(6,0.09457333386)(8,0.07149577886)(10,0.0494672209)(12,0.03541572392)(14,0.02588011138)(16,0.01989355497)(18,0.01577038877)(20,0.0134743331)
        };
        \addlegendentry{DeepRx, tdl\_c}
        \addplot [thick,orange,mark=x,dashed,mark options=solid,mark size=4pt]
        coordinates{
        (0,0.2291197778)(2,0.1848350556)(4,0.1454729444)(6,0.1080377222)(8,0.07989183333)(10,0.05392533333)(12,0.03622377778)(14,0.02370122222)(16,0.01573472222)(18,0.01027572222)(20,0.006853444444)
        };
        \addlegendentry{TradRx, tdl\_c}
        \addplot [thick,Plum,mark=*]
        coordinates{
        (0,0.1543480009)(2,0.1116706654)(4,0.07328022271)(6,0.04114655405)(8,0.01922533289)(10,0.00740161119)(12,0.002598555526)(14,0.0008122777799)(16,0.0002460000105)(18,0.0001301666634)(20,5.03e-05)
        };
        \addlegendentry{DeepRx, tdl\_d}
        \addplot [thick,Plum,mark=*,dashed,mark options=solid]
        coordinates{
        (0,0.1895611667)(2,0.1426593333)(4,0.09959444444)(6,0.06146655556)(8,0.03272011111)(10,0.01466611111)(12,0.005818611111)(14,0.001974555556)(16,0.0005868888889)(18,0.0002397222222)(20,7.52e-05)
        };
        \addlegendentry{TradRx, tdl\_d}
        \addplot [thick,ForestGreen,mark=square*]
        coordinates{
        (0,0.1543922275)(2,0.1111510545)(4,0.07456927747)(6,0.0417637229)(8,0.020136334)(10,0.008187666535)(12,0.002746388782)(14,0.0008083888679)(16,0.0003113888961)(18,5.87e-05)(20,4.07e-05)
        };
        \addlegendentry{DeepRx, tdl\_e}
        \addplot [thick,ForestGreen,mark=square*,dashed,mark options=solid]
        coordinates{
        (0,0.1892956667)(2,0.1418317222)(4,0.1008535)(6,0.06219594444)(8,0.03366872222)(10,0.01556272222)(12,0.006011111111)(14,0.001979777778)(16,0.0006874444444)(18,0.0001438888889)(20,6.4e-05)
        };
        \addlegendentry{TradRx, tdl\_e}
        \end{axis}
    \end{tikzpicture}
    \caption{BER vs. Eb/N0 in Channel Profile - Exp. 2, where DeepRx is trained on tdl\_d (LOS). During test, DeepRx underperforms TradRx for tdl\_a, tdl\_b, and tdl\_c (i.e., NLOS) test channels, and therefore, a \emph{performance drop case} is observed.}
    \label{fig:prelim-channel-drop}
    \end{minipage}
    \begin{minipage}{0.49\textwidth}
    \vspace{0.5cm}
\centering
\begin{tikzpicture}
        \begin{axis}[
            xlabel = Eb/N0 (dB),
            ylabel = BER (log),
            title={Transmitter Speed  - Exp. 1},
            ymin=0.00002,
            ymax=0.2,
            ymode = log,
            xmin=0,
            xmax=20,
            height=\pltheight,
            width=\linewidth,
            legend style={at={(0,0)},
            fill=gray!5,
            fill opacity=0.6,
            text opacity=1,
            draw=none,
            font=\small,
            legend cell align=left,
                anchor=south west,legend columns=1},
            ymajorgrids=true,
            ytick = {0.1,0.01,0.001,0.0001},
            yticklabels = {$10^{-1}$,$10^{-2}$,$10^{-3}$,$10^{-4}$},
            grid style={line width=1pt,draw=gray!50},
        ]
        \addplot [thick,blue,mark=|,mark size=4pt]
        coordinates{
        (0,0.1552288383)(2,0.1104935035)(4,0.07248549908)(6,0.03963744268)(8,0.01738199964)(10,0.006191055756)(12,0.001663388917)(14,0.0004705000028)(16,0.0001960555528)(18,7.11e-05)(20,2.77e-05)
        };
        \addlegendentry{DeepRx, 1m/s}
        \addplot [thick,blue,dashed,mark=|,mark options=solid,mark size=4pt]
        coordinates{
        (0,0.1862243889)(2,0.1374879444)(4,0.09585522222)(6,0.05811183333)(8,0.02942561111)(10,0.0122905)(12,0.003839388889)(14,0.001094333333)(16,0.0003259444444)(18,9.22e-05)(20,2.86e-05)
        };
        \addlegendentry{TradRx, 1m/s}
        \addplot [thick,red,mark=o]
        coordinates{
        (0,0.1556053907)(2,0.1121576652)(4,0.07246116549)(6,0.04093483463)(8,0.01908855513)(10,0.007806944661)(12,0.002732611028)(14,0.0008817221969)(16,0.0002807777782)(18,0.0001110555531)(20,5.09e-05)
        };
        \addlegendentry{DeepRx, 16m/s}
        \addplot [thick,red,mark=o,dashed,mark options=solid]
        coordinates{
        (0,0.1910320556)(2,0.1435973333)(4,0.0993195)(6,0.06206955556)(8,0.03323927778)(10,0.015815)(12,0.006417111111)(14,0.0022545)(16,0.0007338888889)(18,0.0002677222222)(20,0.000106)
        };
        \addlegendentry{TradRx, 16m/s}
        \addplot [thick,ForestGreen,mark=x,mark size=4pt]
        coordinates{
        (0,0.1554559469)(2,0.1117151082)(4,0.0727057755)(6,0.04164522141)(8,0.019952612)(10,0.007635277696)(12,0.002657833276)(14,0.0008198888972)(16,0.0002681111218)(18,0.0001122777758)(20,3.02e-05)
        };
        \addlegendentry{DeepRx, 17m/s}
        \addplot [thick,ForestGreen,mark=x,mark options=solid,dashed,mark size=4pt]
        coordinates{
        (0,0.191244)(2,0.1432714444)(4,0.09973344444)(6,0.06292172222)(8,0.03447677778)(10,0.01567722222)(12,0.006317166667)(14,0.002192444444)(16,0.0007355555556)(18,0.0002683888889)(20,8.28e-05)
        };
        \addlegendentry{TradRx, 17m/s}
        \end{axis}
    \end{tikzpicture}
    \caption{BER vs. Eb/N0 in Transmitter Speed - Exp. 1, where DeepRx is trained on higher speeds of 18, 19, and 20 m/s. During test, DeepRx outperforms TradRx and no \emph{performance drop case} is observed.\\}
    \label{fig:prelim-speed-trained-high}
    \end{minipage}
    \begin{minipage}{0.49\textwidth}
    \vspace{0.5cm}
\centering
\begin{tikzpicture}
        \begin{axis}[
            xlabel = Eb/N0 (dB),
            ylabel = BER (log),
            title={Transmitter Speed - Exp. 2},
            ymin=0.00003,
            ymax=0.2,
            ymode = log,
            xmin=0,
            xmax=20,
            height=\pltheight,
            width=\linewidth,
            legend style={at={(0,0)},
            fill=gray!5,
            fill opacity=0.6,
            text opacity=1,
            draw=none,
            font=\small,
            legend cell align=left,
                anchor=south west,legend columns=1},
            ymajorgrids=true,
            ytick = {0.1,0.01,0.001,0.0001},
            yticklabels = {$10^{-1}$,$10^{-2}$,$10^{-3}$,$10^{-4}$},
            grid style={line width=1pt,draw=gray!50},
        ]
        \addplot [thick,blue,mark=|,mark size=4pt]
        coordinates{
        (0,0.1535911709)(2,0.1110531688)(4,0.07080377638)(6,0.04003277794)(8,0.01847844385)(10,0.007154944353)(12,0.002423889004)(14,0.0006735555362)(16,0.0002163333265)(18,8.26e-05)(20,3.62e-05)
        };
        \addlegendentry{DeepRx, 3m/s}
        \addplot [thick,blue,dashed,mark=|,mark options=solid,mark size=4pt]
        coordinates{
        (0,0.1872949444)(2,0.1404910556)(4,0.09568183333)(6,0.05906122222)(8,0.03080611111)(10,0.01347594444)(12,0.004937388889)(14,0.001455333333)(16,0.000411)(18,0.0001117777778)(20,4.43e-05)
        };
        \addlegendentry{TradRx, 3m/s}
        \addplot [thick,red,mark=o]
        coordinates{
        (0,0.1545031071)(2,0.1118877232)(4,0.0717933327)(6,0.04123122245)(8,0.02046438865)(10,0.007921333425)(12,0.003040944459)(14,0.0009102777694)(16,0.0003660555521)(18,0.0001503888925)(20,8.59e-05)
        };
        \addlegendentry{DeepRx, 4m/s}
        \addplot [thick,red,mark=o,dashed,mark options=solid]
        coordinates{
        (0,0.1877863889)(2,0.1410843333)(4,0.09630661111)(6,0.05983177778)(8,0.03242472222)(10,0.01404188889)(12,0.005475166667)(14,0.001625611111)(16,0.0004977222222)(18,0.0001370555556)(20,5.89e-05)
        };
        \addlegendentry{TradRx, 4m/s}
        \addplot [thick,ForestGreen,mark=x,mark size=4pt]
        coordinates{
        (0,0.1742708385)(2,0.1365091652)(4,0.1037800536)(6,0.0772723332)(8,0.05761677772)(10,0.04395955428)(12,0.03473944589)(14,0.03006866574)(16,0.02697538957)(18,0.02511011064)(20,0.02378283255)
        };
        \addlegendentry{DeepRx, 20m/s}
        \addplot [thick,ForestGreen,mark=x,mark options=solid,dashed,mark size=4pt]
        coordinates{
        (0,0.19157066666666647)(2,0.1435432777777777)(4,0.10021772222222229)(6,0.06280311111111107)(8,0.03451105555555555)(10,0.01633038888888888)(12,0.006392166666666668)(14,0.0024174444444444426)(16,0.0008306666666666662)(18,0.00031716666666666677)(20,0.00014200000000000047)
        };
        \addlegendentry{TradRx, 20m/s}
        \end{axis}
    \end{tikzpicture}
    \caption{BER vs. Eb/N0 in Transmitter Speed - Exp. 2, where DeepRx is trained on lower speeds of 0, 1, and 2 m/s. During test, DeepRx underperforms TradRx in speeds 4 m/s and larger, which shows a \emph{performance drop case}.}
    \label{fig:prelim-speed-trained-low}
    \end{minipage}
    \begin{minipage}{0.49\textwidth}
    \vspace{0.5cm}
\centering
\begin{tikzpicture}
        \begin{axis}[
            xlabel = Eb/N0 (dB),
            ylabel = BER (log),
            title={Delay Spread - Exp. 1},
            ymax=0.25,
            ymode = log,
            xmin=0,
            xmax=20,
            height=\pltheight,
            width=\linewidth,
            legend style={at={(0,0)},
            fill=gray!5,
            fill opacity=0.6,
            text opacity=1,
            draw=none,
            font=\small,
            legend cell align=left,
                anchor=south west,legend columns=1},
            ymajorgrids=true,
            ytick = {0.1,0.01,0.001},
            yticklabels = {$10^{-1}$,$10^{-2}$,$10^{-3}$},
            grid style={line width=1pt,draw=gray!50},
        ]
        \addplot [thick,blue, mark=|,mark size=4pt]
        coordinates{ 
        (0,0.1641860604)(2,0.120535776)(4,0.08071538806)(6,0.05244800076)(8,0.02764316648)(10,0.0167778898)(12,0.008285500109)(14,0.005168888718)(16,0.003132055514)(18,0.001674777828)(20,0.001256500022)
        };
        \addlegendentry{DeepRx, 10 ns}
        \addplot [thick,dashed,blue,mark=|,mark options=solid,mark size=4pt]
        coordinates{ 
        (0,0.1991082778)(2,0.1517038333)(4,0.1072696667)(6,0.07341144444)(8,0.04163238889)(10,0.02543055556)(12,0.01267461111)(14,0.007519277778)(16,0.004485277778)(18,0.002308777778)(20,0.001682722222)
        };
        \addlegendentry{TradRx, 10 ns}
        \addplot [thick,red,mark=o]
        coordinates{
        (0,0.1734752804)(2,0.1283884495)(4,0.098267667)(6,0.06567060947)(8,0.03896927834)(10,0.02242261171)(12,0.01470305584)(14,0.01005133335)(16,0.005783166736)(18,0.003776000114)(20,0.002117055468)
        };
        \addlegendentry{DeepRx, 50 ns}
        \addplot [thick,red,dashed,mark=o,mark options=solid]
        coordinates{
       (0,0.2088373889)(2,0.1599031667)(4,0.1258997222)(6,0.08754455556)(8,0.05452755556)(10,0.03254144444)(12,0.020984)(14,0.01398561111)(16,0.008076333333)(18,0.005226777778)(20,0.002863444444)
        };
        \addlegendentry{TradRx, 50 ns}
        \addplot [thick,ForestGreen,mark=x,mark size=4pt]
        coordinates{
        (0,0.1765721738)(2,0.1367129385)(4,0.0959315002)(6,0.06749510765)(8,0.0465051122)(10,0.03100199997)(12,0.01718166657)(14,0.01050250046)(16,0.008667611517)(18,0.004449444357)(20,0.002955166623)
        };
        \addlegendentry{DeepRx, 80 ns}
        \addplot [thick,ForestGreen,mark=x, mark options=solid, dashed,mark size=4pt]
        coordinates{
        (0,0.2120328333)(2,0.1684706667)(4,0.1230889444)(6,0.08942755556)(8,0.06296477778)(10,0.04242488889)(12,0.02392155556)(14,0.01453577778)(16,0.01173861111)(18,0.006183722222)(20,0.004052333333)
        };
        \addlegendentry{TradRx, 80 ns}
        \end{axis}
    \end{tikzpicture}
    \caption{BER vs. Eb/N0 in Delay Spread - Exp. 1, where DeepRx is trained on higher delay spreads of 400, 450, and 500 ns. During test, DeepRx outperforms TradRx and no \emph{performance drop case} is observed.}
    \label{fig:prelim-delay-trained-high}
    \end{minipage}
    \begin{minipage}{0.49\textwidth}
    \vspace{0.5cm}
\centering
\begin{tikzpicture}
        \begin{axis}[
            xlabel = Eb/N0 (dB),
            ylabel = BER (log),
            title={Delay Spread - Exp. 2},
            ymin=0.003,
            ymax=0.3,
            ymode = log,
            xmin=0,
            xmax=20,
            height=\pltheight,
            width=\linewidth,
            legend style={at={(0,0)},
            fill=gray!5,
            fill opacity=0.6,
            text opacity=1,
            draw=none,
            font=\small,
            legend cell align=left,
                anchor=south west,legend columns=1},
            ymajorgrids=true,
             ytick = {0.1,0.01,0.001},
            yticklabels = {$10^{-1}$,$10^{-2}$,$10^{-3}$},
            grid style={line width=1pt,draw=gray!50},
        ]
        \addplot [thick,blue,mark=|,mark size=4pt]
        coordinates{ 
        (0,0.1831858903)(2,0.1383752823)(4,0.1000140533)(6,0.07448133081)(8,0.04312538728)(10,0.03029255569)(12,0.01804799959)(14,0.01147027779)(16,0.007259944454)(18,0.004493888933)(20,0.003116277745)
        };
        \addlegendentry{DeepRx, 100 ns}
        \addplot [thick,dashed,blue,mark=|,mark options=solid,mark size=4pt]
        coordinates{ 
        (0,0.2191907778)(2,0.1702879444)(4,0.1269691111)(6,0.09656261111)(8,0.05821911111)(10,0.04121166667)(12,0.02481427778)(14,0.01557188889)(16,0.009853333333)(18,0.006003333333)(20,0.004016833333)
        };
        \addlegendentry{TradRx, 100 ns}
        \addplot [thick,red,mark=o]
        coordinates{
        (0,0.1894153953)(2,0.1401918381)(4,0.1109232232)(6,0.08265060931)(8,0.05355966836)(10,0.03871616721)(12,0.02518822253)(14,0.01503455546)(16,0.01139233354)(18,0.007172999904)(20,0.005268166773)
        };
        \addlegendentry{DeepRx, 200 ns}
        \addplot [thick,red,dashed,mark=o,mark options=solid]
        coordinates{
        (0,0.2231191667)(2,0.1694836667)(4,0.1354577222)(6,0.1019895556)(8,0.06678994444)(10,0.04774438889)(12,0.03024444444)(14,0.01772966667)(16,0.01235716667)(18,0.007339666667)(20,0.004807222222)
        };
        \addlegendentry{TradRx, 200 ns}
        \addplot [thick,ForestGreen,mark=x,mark size=4pt]
        coordinates{
        (0,0.1976507157)(2,0.1647330523)(4,0.1297672242)(6,0.1018351093)(8,0.07736944407)(10,0.05892549828)(12,0.04741316661)(14,0.0362017788)(16,0.03270938993)(18,0.02672172152)(20,0.0229691118)
        };
        \addlegendentry{DeepRx, 400 ns}
        \addplot [thick,ForestGreen,mark=x,dashed,mark options=solid,mark size=4pt]
        coordinates{
        (0,0.2204461667)(2,0.1809826111)(4,0.1395457222)(6,0.1044756111)(8,0.07315488889)(10,0.04876238889)(12,0.03403594444)(14,0.02134811111)(16,0.01478372222)(18,0.008236666667)(20,0.005022888889)
        };
        \addlegendentry{TradRx, 400 ns}
        \end{axis}
    \end{tikzpicture}
    \caption{BER vs. Eb/N0 in Delay Spread - Exp. 2, where DeepRx is trained on lower delay spreads of 10, 50, and 80 ns. During test, DeepRx underperforms TradRx in 400 ns test set which shows a \emph{performance drop case}.}
    \label{fig:prelim-delay-trained-low}
    \end{minipage}
\end{figure*}

The superior BER performance of DeepRx compared to the traditional receiver is previously evaluated in the original paper by Nokia Bell Labs~\cite{deeprx}. National Instruments also shows DeepRx BER comparisons against the traditional receiver in a real-time system prototype built using their universal software radio peripherals (USRPs)~\cite{ni-white-deeprx}. However, in both works the training and test datasets have the same configurations. Here, we attempt to identify corner cases where DeepRx BER increases above the traditional receiver BER (a.k.a., \emph{performance drop cases}). While we limit our studies to DeepRx as a widely used 5G receiver, our core method can be deployed to any general AI-native receiver.

In our preliminary experiments, we vary three parameters--channel profile, transmitter speed, and delay spread--between the training and test sets, and study their impact on DeepRx BER. In all training and test datasets, data modulation scheme is set to 16QAM, and AWGN is added such that the ratio of energy per bit to the spectral noise density (Eb/N0) is in range 0 to 20 dB with steps of 2 dB. Each training and test set contains 5000 and 500 uplink 5G radio frames, respectively, per Eb/N0 level, and per combination of channel profile, transmitter speed, and delay spread. In each training run, the DeepRx model is fully trained for $\sim$20 epochs. To measure TradRx BER and DeepRx BER in different Eb/N0 levels, the corresponding test set is passed through the TradRx, and the trained DeepRx model, respectively, and BER versus Eb/N0 is plotted in Figs.~\ref{fig:prelim-channel-pass}-\ref{fig:prelim-delay-trained-low}. It is expected that DeepRx BER is lower than TradRx BER in all Eb/N0 levels (i.e., DeepRx outperforms TradRx), otherwise that specific training/test configuration is flagged as a \emph{performance drop case}. We perform 6 training experiments with different training and test set parameters as summarized in Table~\ref{tab:exp_description}.


\begin{table}[t]
    \centering
    \renewcommand{\arraystretch}{1.2} 
    \resizebox{1.0\linewidth}{!}{
    \begin{tabular}{|c|l||c|c|c|}
        \hline
        \multicolumn{2}{|c||}{ } & Channel Profile & Tx Speed & Delay Spread \\
        \multicolumn{2}{|c||}{} & Exp.~2 & Exp.~2 & Exp.~2 \\
        \hline
        \multirow{3}{*}{\rotatebox{90}{Training}} 
            & Channel Profile & tdl\_d & tdl\_d & tdl\_b \\
            & Tx Speed (m/s) & 10 & 0, 1, 2 & 2 \\
            & Delay Spread (ns) & 400 & 400 & 10, 50, 80 \\
        \hline
        \multirow{4}{*}{\rotatebox{90}{Test}} 
            & Channel Profile & tdl\_a, tdl\_b, tdl\_c & tdl\_d & tdl\_b \\
            & Tx Speed (m/s) & 10 & $\geq$ 4 & 2 \\
            & Delay Spread (ns) & 400 & 400 & $>$ 200 \\
        \hline
    \end{tabular}
    }
    \caption{Training and test configurations that lead to DeepRx underperforming TradRx.}
    \label{tab:drop}
\end{table}

Figures~\ref{fig:prelim-channel-pass}-\ref{fig:prelim-delay-trained-low}, show that DeepRx might provide higher BER compared to TradRx in three different cases: (i) change in the channel profile: if DeepRx is trained on a LOS channel such as tdl\_d and deployed in NLOS channel profiles such as tdl\_a, tdl\_b, and tdl\_c. (ii) change in the transmitter speed: if DeepRx is trained on a specific speed range such as speeds 0, 1, and 2 m/s and is tested on speeds that are higher than the training speeds by at least 2 m/s. (iii) change in the delay spread: if DeepRx is trained on low delay spreads such as 10, 50, and 80 ns and tested on higher delay spreads such as 400 ns. It should be noted that DeepRx performance drop cases are not limited to the above three configurations, however, these explored cases are summarized in Table~\ref{tab:drop} as example configurations that lead to DeepRx performance drop. 

Next, we introduce VERITAS that automatically detects performance drop cases during the deployment of AI-native receiver.
\section{VERITAS for Verifying the BER Performance of AI-native Receivers}\label{sec:method}
In this section, we describe VERITAS as a framework for verifying the performance of AI-native receiver. We describe the overview of VERITAS and the interactions between its different components in Section~\ref{sec:method-system}. We provide the details of the \emph{Monitor}, and the \emph{Performance Comparator} in Sections~\ref{sec:method-monitor} and~\ref{sec:method-comparator}, respectively.

\vspace{-4mm}
\subsection{VERITAS System Overview}\label{sec:method-system} 
As shown in Fig.~\ref{fig:overview}, VERITAS, is placed in parallel to the AI-native receiver and consists of three different components: the \emph{Monitor}, the \emph{Performance Comparator}, and the TradRx (introduced earlier in Section~\ref{sec:prelim-pipelines}). Among these components, the Monitor is the only component that is continuously active, while the other components are triggered based on certain conditions. The Monitor that is an NN cascaded with an OOD detection algorithm is pretrained on the same training set as the AI-native receiver. Therefore, the specific channel profiles, transmitter speeds, and delay spreads covered in the training set are considered ID data for the Monitor. The Monitor constantly observes the wireless channel by processing received pilots and flags potential OOD pilots as change in the wireless channel. Following this change detection, one could retrain the AI-native receiver to adapt it to the new wireless channel, however, not all changes might cause performance drop for the AI-native receiver (as observed in Figs.~\ref{fig:prelim-channel-pass}, \ref{fig:prelim-speed-trained-high}, and \ref{fig:prelim-delay-trained-high}), and retraining might be unnecessary. To avoid unnecessary retraining, at this point the Performance Comparator is activated and TradRx is triggered to run as the reference point in parallel to the AI-native receiver. The Performance Comparator compares bit probabilities generated by TradRx and the AI-native receiver, and determines the underperforming receiver. If the AI-native receiver outperforms TradRx, no retraining is required. In this case the TradRx and the Performance Comparator are deactivated and the Monitor continues observing the wireless channel to detect future potential changes. If the AI-native receiver underperforms TradRx, a retraining process is initiated to adapt the AI-native receiver to the new wireless channel. If a retraining process is initiated for the AI-native receiver, the Monitor needs to be retrained as well to update its set of ID classes to be able to continue detecting further changes in the wireless channel.

\vspace{-3mm}

\subsection{Wireless Channel Change Detector: Monitor}\label{sec:method-monitor}
\begin{figure}[t!!!]
    \centering
    \includegraphics[width=\linewidth]{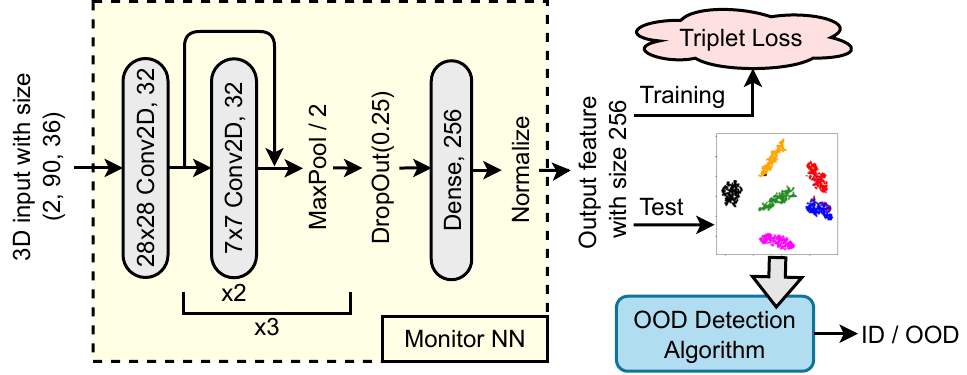}
    \caption{The Monitor with an NN with $\sim$712k parameters that is trained with triplet loss function. During deployment the test features at the output of NN are fed to an OOD detection algorithm to give an ID or OOD decision for each input.}
    \label{fig:monitor-arch}
\end{figure}

The job of the Monitor is to observe the wireless channel and detect potential changes in the channel profile, transmitter speed, or delay spread, which can be formulated as an OOD detection task. Among different varieties of OOD detection algorithms~\cite{ood-survey,open-ood}, we adopt a feature-based OOD detection method from the post-hoc category, due to implementation simplicity and efficiency. We design a custom NN that generates vectorized features and train it on the training set containing the ID data classes, without getting exposed to any representation of OOD data during training. During deployment, the generated output features are fed to a novel distance-based OOD detection algorithm, to make an ID or OOD decision for each input. An overview of the Monitor is shown in Fig.~\ref{fig:monitor-arch}.

In the following, we describe the input and output of the Monitor NN in Section~\ref{sec:method-monitor-io}, the Monitor NN architecture along with the training and test processes in Section~\ref{sec:method-monitor-arch}, and the proposed OOD detection algorithm for detecting wireless channel changes in Section~\ref{sec:method-monitor-ood}.

\subsubsection{Input and Output Structure}\label{sec:method-monitor-io}
Here, we explain how Monitor input is prepared, and what output the Monitor provides. 

\begin{figure}[t!!!]
    \centering
    \includegraphics[width=\linewidth]{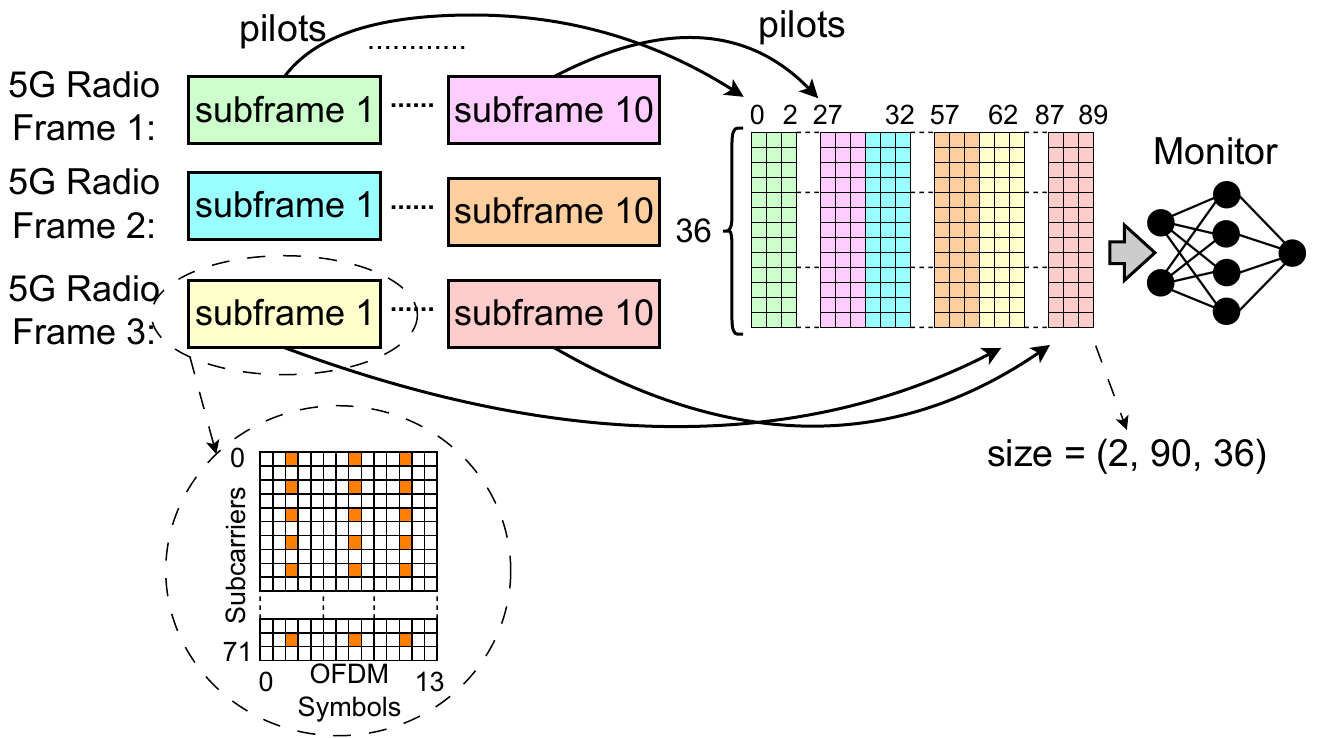}
    \caption{To construct each input for the Monitor NN, received pilots in three 5G radio frames are concatenated to form 3D input tensors with dimensions (2, 90, 36).}
    \label{fig:monitor-input}
\end{figure}

\noindent \textbf{Input.} The wireless channel change detection happens through processing the frequency domain representation of the received 5G pilots. Specifically, we create a 3D matrix using pilots in 3 consecutive received frames. Constricting the input of the Monitor does not impose additional signal processing steps to the system, as the received frequency domain pilots are already prepared as an input component to DeepRx~\cite{deeprx}. Depending on the selected pilot pattern, the number of pilot columns will be different which leads to different input sizes for the Monitor. In our selected pilot pattern (see the bottom of Fig.~\ref{fig:monitor-input}) each OFDMA subframe consists of complex-valued pilots with dimensions 36 and 3 along the frequency and time axes, respectively. We take pilot matrices from all the 10 subframes in 3 consecutive 5G radio frames and concatenate them along the time axis. We separate the real and imaginary parts of the pilots and form a matrix (tensor) with size (2, 90, 36) that is the input to the Monitor NN. The process of preparing inputs for the Monitor using 3 consecutive 5G radio frames each comprising 10 subframes is shown in Fig.~\ref{fig:monitor-input}.

\noindent \textbf{Output.} The output of the Monitor is a binary decision (ID or OOD) per input tensor.

\subsubsection{NN Architecture and Training Process}\label{sec:method-monitor-arch}
For the Monitor NN architecture, we design a custom convolutional NN with residual blocks, consisting of convolutional, maxpooling, dense (or fully connected), and dropout layers, as shown in Fig.~\ref{fig:monitor-arch}. We design the last layer of the NN to be a dense layer with output size $I$~=~256, and normalize its output, $x$, as in~\eqref{eq:normalize}, before sending it out of the NN.
\begin{equation}\label{eq:normalize}
    y = \frac{x}{\max(|x|)}
\end{equation}
Here, $\max(x)$ returns the maximum value among all the elements in vector $x$, and $|\cdot|$ is the absolute operator.
We note that $y$ is a vector of $I$ elements, however, we do not denote its vector indexing in this paper for the sake of simplicity. 

We train the Monitor NN with triplet loss function~\cite{triplet-loss} to create clustered embeddings from input data. Triplet loss function operates on three input vectors: anchor feature and positive feature belonging to the same class, and negative feature belonging to a different class. Based on this, triplet loss function on a mini-batch comprising $\mathcal{N}$ triples of input samples is defined as~\eqref{eq:triplet_loss}.
\begin{equation}\label{eq:triplet_loss}
    \mathcal{L} = \sum_{i=0}^{\mathcal{N}-1} \max\{\norm{x_a^{(i)}-x_p^{(i)}}_2^2 - \norm{x_a^{(i)}-x_n^{(i)}}_2^2 + \alpha, 0\}
\end{equation}
In~\eqref{eq:triplet_loss}, $x_a^{(i)}$, $x_p^{(i)}$, and $x_n^{(i)}$ denote $i^{\text{th}}$ anchor, positive, and negative features, respectively. $\norm{z_1-z_2}_2$ denotes Euclidean distance between any given two vectors $z_1$ and $z_2$, and $\max(w_1,w_2)$ returns the largest value among $w_1$ and $w_2$ scalars. $\alpha$ is the triplet loss margin that we keep as the default value 1. In this way, the objective of training is to minimize $\mathcal{L}$ that is a sum over $\mathcal{N}$ loss components in each mini-batch.

The implication behind \eqref{eq:triplet_loss} is to map the positive features as close as possible to the anchor features, and to map the negative features as far as possible from the anchor feature in the feature space, and hence, to form distinct clusters for all the training set classes.

\subsubsection{OOD Detection Algorithm}\label{sec:method-monitor-ood}
To detect a change in the wireless channel that is equivalent to identifying 5G pilot matrices as OOD, we propose a non-parametric OOD detection algorithm that processes the output features of the Monitor NN, $y$ vectors. As the NN is trained with triplet loss function, our algorithm relies on the assumption that features generated from inputs belonging to the same ID class fall in the same cluster, and features generated from inputs belonging to different ID classes form distinct clusters. The proposed algorithm flags each test feature as ID or OOD by comparing distances of the unknown test feature and known ID features to the center of the ID clusters. As distance calculations for the complete set of features within an ID cluster is computationally very costly, we limit our distance calculations to the $K$ ID features that are closest to the test feature and are actually the most critical among all the ID features for ID or OOD decision. Based on this, we take advantage of the KNN~\cite{knn} algorithm to find the $K$ nearest neighbors of each test feature. In the following, we describe the processes of characterizing ID clusters and fitting a KNN model to them that happen pre-deployment and ID/OOD decision making for each test sample that happens at the deployment phase.

\noindent \textbf{Characterizing ID Clusters:} The multidimensional ID clusters are created by passing the ID classes (i.e., the training set) through the fully trained Monitor NN, in the pre-deployment phase. In a training dataset with $J$ ID classes indexed with $j=0,...,J-1$, where each ID class has population $N_{j}$, output features of the Monitor NN are denoted as $y_j^{(n)}$ with $n=0,...,N_{j}-1$. Each ID cluster needs to be characterized with a center, $c_j$, that is a vector of size $I$ and a radius, $r_j$, that is a scalar. In this case, we collect all output vectors, $y_j^{(n)}$, belonging to each ID class $j$, and calculate a center, $c_j$, for each ID cluster using~\eqref{eq:center}.
\begin{equation}\label{eq:center}
    c_j = \frac{1}{N_{j}}\sum_{n=0}^{N_{j}-1} y_j^{(n)}, \ \ \ \   j=0,1,...,J-1
\end{equation}
Equation~\eqref{eq:center} simply calculates an $I$-dimensional mean for all the $I$-dimensional features in each ID class. To calculate cluster radius, $r_j$, associated with each ID class $j$ in the training set, we calculate the Euclidean distance of all training features, $y_j^{(n)}$, from $c_j$, and sort them in the ascending order. We discard the last $1-\lambda$ portion of the sorted list and keep the first $\lambda$ portion. The portion that is discarded is associated with those features that are far away from the cluster center, $c_j$. We pick the last (i.e., largest) value of the remaining list as the cluster radius, $r_j$. We set $\lambda$ to a large value such as 95\%, so that the distances of 95\% of the ID features to the cluster center, $c_j$, are smaller than the cluster radius, $r_j$ (i.e., 95\% of ID features fall inside their respective clusters). The aforementioned steps for characterizing each cluster by a center and a radius are summarized in Algorithm~\ref{alg:cluster-characterizing}.

\begin{algorithm}[t]
    \caption{Characterizing ID Clusters}
    \begin{algorithmic}[1]
        \STATE \textbf{Inputs:} Trained Monitor NN, Training set containing all ID classes \\
        \STATE Set $\lambda$ as 95\%
        \STATE Pass the training set through the trained Monitor NN and collect the ID features, $y_j^{(n)}$ vectors
        \STATE center\_list , radius\_list  = [ ] , [ ] \\
        \FOR{class $j$ in ID\_class\_list} 
        \STATE Calculate cluster center $c_j$ using \eqref{eq:center} \\
        \STATE center\_list.append($c_j$)\\
        \STATE distance\_list = [ ] \\
            \FOR{$y_j^{(n)}$ in class $j$}
                \STATE distance\_list.append($\norm{y_j^{(n)}-c_j}_2$)
            \ENDFOR
            \STATE Sort the distance\_list in ascending order \\
            \STATE distance\_list = distance\_list [start  : $\lambda \times$end]
            \STATE $r_j$ = distance\_list [end]
            \STATE radius\_list.append($r_j$)\\
        \ENDFOR
        \STATE \textbf{Outputs:} center\_list, radius\_list
    \end{algorithmic}
    \label{alg:cluster-characterizing}
\end{algorithm}

\noindent \textbf{Fitting the KNN Model.} As the final step in the pre-deployment phase, we combine all the ID features from different ID classes in a single set and fit a KNN model to them, using \texttt{NearestNeighbors} class from \texttt{sklearn.neighbors} library.

\begin{algorithm}[t]
    \caption{OOD Detection}
    \begin{algorithmic}[1]
        \STATE\textbf{Inputs:} center\_list, radius\_list, test feature $y_{\text{test}}$ \\
        \vspace{1mm}
        \STATE Return $K$ nearest neighbors as n\_list = $\bigcup_{k=0}^{K-1} \text{neighbor}_k$ and find their corresponding cluster centers using center\_list and record them as c\_list = $\bigcup_{k=0}^{K-1} c_j^{(k)}$\\
        vote\_list = [ ]
        \FOR{($\text{neighbor}_k$ , $c_j^{(k)})$ in \textbf{zip} (n\_list , c\_list)}
        \vspace{2mm}
            \STATE $d_k = \norm{\text{neighbor}_k-c_j^{(k)}}_2$ \quad , \quad $d_y = \norm{y_{\text{test}}-c_j^{(k)}}_2$ \\
            \vspace{2mm}
            \STATE $v_k$ = ID \textbf{if} ($d_y \leq d_k$ \textbf{and} $d_y \leq r_j$) \textbf{else} OOD
        \ENDFOR
        \STATE $v_{\text{final}}$ = ID \textbf{if} ID $\in$ vote\_list \textbf{else} OOD
        \STATE \textbf{Output:} $v_{\text{final}}$
    \end{algorithmic}
    \label{alg:ood}
\end{algorithm}

\noindent \textbf{Making ID/OOD Decision for Each Test Sample:}
At the deployment phase, the Monitor is tested on input samples that might belong to an ID class or might be OOD. For each output vector $y_{\text{test}}$ generated using each test input, we find its $K$ nearest neighbors, each denoted as $\text{neighbor}_k$ with $k=0,..., K-1$. Obviously, each of these neighbors belong to one of the ID classes. For each $\text{neighbor}_k$ belonging to the ID class $j$, we find the Euclidean distance of $\text{neighbor}_k$ to its corresponding cluster center $c_j$, and denote it as $d_k$. We also calculate the Euclidean distance of $y_{\text{test}}$ from $\text{neighbor}_k$'s associated cluster center, $c_j$, and denote it as $d_y$. After this, we take a vote from each $\text{neighbor}_k$ belonging to ID class $j$, determining whether the test output feature $y_{\text{test}}$ belongs to class $j$ or not. We check 2 criteria to get the vote of $\text{neighbor}_k$ denoted as $v_k$, as in~\eqref{eq:vote}. 
\begin{equation}\label{eq:vote}
    v_k = 
    \begin{cases}
       \text{ID} &\text{if } d_y \leq d_k \text{ and } d_y \leq r_j  \\
       \text{OOD} &\text{otherwise}\\
    \end{cases}
\end{equation}
As shown in~\eqref{eq:vote}, a feature, $y_{\text{test}}$, is voted as ID by $\text{neighbor}_k$ if the Euclidean distance of that feature to the cluster center of the neighbor is smaller or equal to the distance of the neighbor to its cluster center, and the distance of the feature from the cluster center of the neighbor is smaller than the cluster radius. Otherwise, the sample is voted as OOD by that neighbor.

We derive a final joint vote, $v_{\text{final}}$, for each $y_{\text{test}}$ using the votes from its $K$ neighbors as in~\eqref{eq:finalvote}.
\begin{equation}\label{eq:finalvote}
    v_{\text{final}} = 
    \begin{cases}
       \text{ID} &\text{if } \text{any }v_k = \text{ID},\ \ \ k=0,...,K-1   \\
       \text{OOD} &\text{otherwise}\\
    \end{cases}
\end{equation}
Basically, we identify each sample, $y_\text{test}$, as OOD if none of its nearest neighbors vote it to be ID with respect to their own ID clusters. The steps to identify each test sample as ID or OOD is summarized in Algorithm~\ref{alg:ood}.

The highlights of the proposed OOD detection algorithm with respect to the state-of-the-art proposed in~\cite{ood-knn} and~\cite{RF-fingerprinting-lora} are as follows:

\begin{itemize}
    \item Our proposed OOD detection algorithm performs well in sparse clusters with lower density in the center and higher density around the edges, due to being dependent on the ID cluster center and radius instead of the distance of the test sample from its nearest ID neighbors.
    \item Our proposed algorithm utilizes triplet loss (Section~\ref{sec:method-monitor-arch}) instead of supervised contrastive loss~\cite{SupConLoss} used in~\cite{ood-knn} during training that allows for good true OOD detection rate as well as low false positive rate.
    \item We use a custom NN architecture for OOD detection and show that non-parametric feature-based OOD detection is applicable to 5G wireless data besides the benchmark image datasets of CIFAR, SVHN, etc. demonstrated in~\cite{ood-knn}.
\end{itemize}

If the Monitor detects a change in the wireless channel, the Performance Comparator is activated that is described next.

\vspace{-3mm}
\subsection{Performance Comparator}\label{sec:method-comparator}
The job of the Performance Comparator in VERITAS is to decide if the AI-native receiver (e.g., DeepRx) needs to be retrained. It compares the bit probabilities generated by TradRx and DeepRx to determine the receiver with higher BER, and initiates a retraining process only if DeepRx yields higher BER. Obviously, this comparison happens using predicted bit probabilities without actual BER calculation or access to true bit labels. The Performance Comparator is able to operate on probabilities associated with encoded as well as uncoded bits, which obviates the need to include a costly decoding operation within VERITAS. 
 
As soon as the Performance Comparator is activated by the Monitor, it triggers TradRx and runs it in parallel to DeepRx to decode the same received 5G radio frames for a specific time duration. We collect the softbits (i.e., LLRs) out of both receivers in this time duration, and convert them to bit probabilities. Bit probability $P$ is calculated using its corresponding LLR through \eqref{eq:llr}.
\begin{equation}\label{eq:llr}
    P = \frac{1}{1+e^{\text{LLR}}}
\end{equation}
In the hard decoding method, if $P$ is less than or equal to 0.5 the bit is translated to logical `0', and if $P$ is greater than 0.5 the bit is translated to logical `1'. Since the same received 5G radio frames are passed through DeepRx and TradRx, ideally the same bit probabilities should be generated by both receivers. However, we discover that in practice this is not the case. We find bit probabilities of TradRx and DeepRx to be different for the same radio frames, and even more, we find these probabilities to be related to each receiver's BER. Based on our findings, we derive an empirical method that is inspired by histogram binning approach~\cite{calibration}, which is a basic scheme for calibrating NN predicted probabilities. We note that we only leverage the ``binning'' concept without performing any sort of calibration or modification on DeepRx bit probabilities. We detail the proposed method in the following.

As each bit probability is a value between 0 and 1, we break the range 0 to 1 into 10 non-overlapping bins indexed with $b$, as in \eqref{eq:bin}.\vspace{-2mm}
\begin{equation}\label{eq:bin}
    \text{bin}_b : 
    \begin{cases}
       [\frac{b-1}{10},\frac{b}{10}) &b=1,2,3,...,9  \\
       [\frac{b-1}{10},\frac{b}{10}] &b=10\\
    \end{cases}
\end{equation}

\begin{figure}[t!!!]
    \centering
    \includegraphics[width=\linewidth]{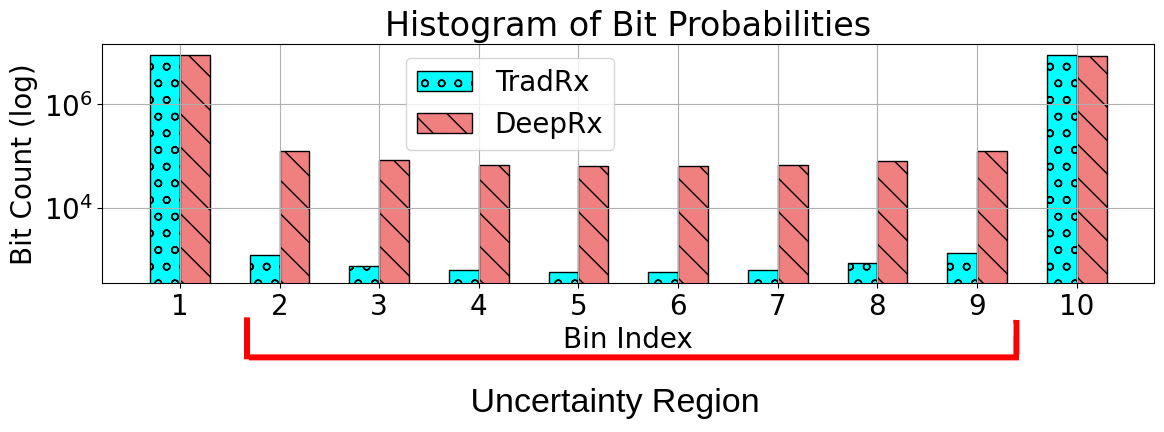}
    \caption{Histogram of 4.5 million bit probabilities generated by passing the same test set through TradRx and DeepRx.}
    \label{fig:histogram1}
\end{figure}
Next, we categorize all the output bits from TradRx and DeepRx into these bins, based on their probability values, and count the number of bits in each bin. The histogram created for 4.5 million output bits of DeepRx and TradRx is shown in Fig.~\ref{fig:histogram1}. In such a histogram, $\text{bin}_1$ and $\text{bin}_{10}$ represent the most certain predictions for logical bits `0' and `1', respectively. On the other hand, $\text{bin}_2$ to $\text{bin}_9$ represent less certain predictions. We refer to these lower probability bins as the \emph{uncertainty region}. We sum the bit counts in the uncertainty region and refer to them as $\mathcal{U}_\text{DeepRx}$ and $\mathcal{U}_\text{TradRx}$ for DeepRx and TradRx, respectively. The histogram bars in Fig.~\ref{fig:histogram1} are plotted for an example 5G radio frame dataset with transmitter speed set as 20 m/s and Eb/N0 as 20 dB (the same dataset studied in Experiment 4 in Table~\ref{tab:exp_description}). For this dataset, DeepRx provides a higher BER compared to TradRx, and accordingly, in Fig.~\ref{fig:histogram1}, DeepRx shows higher bit counts in the uncertainty region compared to TradRx ($\mathcal{U}_\text{DeepRx}$~=~680k vs. $\mathcal{U}_\text{TradRx}$~=~6k). Statistical analysis over the whole test set of 500 5G radio frames (i.e., 4.5 million bits) in different test speeds and different Eb/N0 levels, verifies the same relation between bit probabilities and bit errors: The receiver with the larger $\mathcal{U}$ (i.e., taller histogram bars in the uncertainty region) yields higher BER. However, to comply with this observation, we require to run TradRx and DeepRx in parallel for 500 5G radio frames and collect 4.5 million bits, only to determine the underperforming receiver and potentially trigger retraining. This imposes significant elapsed time to VERITAS and might cause many cyclic redundancy check (CRC) fails before determining the underperforming receiver, which in turn reduces the overall communication throughput. The important question is \emph{What is the smallest bit population that follows the observed rule regarding the relation of bit probabilities and receiver BER?}
We evaluate the Performance Comparator and answer this question in Section~\ref{sec:eval-comparator}.

The described Performance Comparator algorithm is shown in Algorithm~\ref{alg:comparator}.

\begin{algorithm}[t]
    \caption{Performance Comparator}
    \begin{algorithmic}[1]
        \STATE\textbf{Inputs:} $\text{LLR}_{\text{DeepRx}}$, $\text{LLR}_{\text{TradRx}}$ \\
        \STATE Convert $\text{LLR}_{\text{DeepRx}}$ and $\text{LLR}_{\text{TradRx}}$ to $\text{P}_{\text{DeepRx}}$ and $\text{P}_{\text{TradRx}}$, respectively, using \eqref{eq:llr} \\
        \STATE Create histogram bins for both $P$s using \eqref{eq:bin} \\
        \STATE Calculate $\mathcal{U}_\text{DeepRx}$ as sum of the bit count in $\text{bin}_2$ to $\text{bin}_9$, for DeepRx outputs \\
        \STATE Calculate $\mathcal{U}_\text{TradRx}$ as sum of the bit count in uncertainty region ($\text{bin}_2$ to $\text{bin}_9$), for TradRx outputs \\
        \STATE Retraining = not needed \textbf{if} $\mathcal{U}_\text{DeepRx} \leq \mathcal{U}_\text{TradRx}$ \textbf{else} needed
        \STATE \textbf{Output:} Retraining
    \end{algorithmic}
    \label{alg:comparator}
\end{algorithm}

\vspace{-5mm}
\section{Evaluations}
In this section, we evaluate the performance of different components in VERITAS with respect to changes in the channel profile, the transmitter speed, and the delay spread. We evaluate the Monitor and the Performance Comparator in Sections~\ref{sec:eval-monitor} and \ref{sec:eval-comparator}, respectively.

\vspace{-3mm}
\subsection{Wireless Channel Change Detector: Monitor}\label{sec:eval-monitor}
We train three Monitor NNs with triplet loss function, test the trained NNs and calculate and visualize the results in the following forms:

\noindent (i) \textbf{2D Projection of Features.} We test each trained Monitor NN on all the ID and OOD data in the test set to get the 256-dimensional features. we use t-SNE~\cite{tsne} to reduce feature dimensions to 2, and plot them as scatter plots. We note that t-SNE is used only for visualization, and Algorithms~\ref{alg:cluster-characterizing} and~\ref{alg:ood} operate on the 256-dimensional features, without any dimension reduction. 

\noindent (ii) \textbf{OOD Detection Rate for Different $K$ Values.} 
To numerically evaluate how separable the ID and OOD classes are, we pass the training set containing the ID classes through the trained Monitor NN and record the ID features. We characterize each 256-dimensional ID cluster by a 256-dimensional center $c_j$, and a scalar radius $r_j$ through Algorithm~\ref{alg:cluster-characterizing}. We run the OOD detection algorithm (i.e., Algorithm~\ref{alg:ood}) with $\lambda$~=~0.95 and nearest neighbor $K$ values of 5, 10, and 15, on the unseen test set that contains ID and OOD classes. For each test class, each $K$, and each Eb/N0 level, we calculate OOD detection rate as the number of test feature vectors detected as OOD divided by the total number of test feature vector.

\noindent (iii) \textbf{Sensitivity of Algorithm~\ref{alg:ood} to $\lambda$.} $\lambda$ determines the radii $r_j$ of ID clusters which are inputs to Algorithm~\ref{alg:ood}. We study the sensitivity of Algorithm~\ref{alg:ood} to $\lambda$ by measuring OOD detection rate while varying $\lambda$ in range 0.5 to 1.0, with $K$ = 15, at Eb/N0 levels 0, 10, and 20 dB.

We describe the experiments and results for detecting a change in the channel profile, transmitter speed, and delay spread in Sections~\ref{sec:eval-monitor-channel},~\ref{sec:eval-monitor-speed}, and~\ref{sec:eval-monitor-delay}, respectively.


\subsubsection{Detecting a Change in the Channel Profile}\label{sec:eval-monitor-channel}
\begin{figure*}[t!!!]
    \centering
    \includegraphics[width=0.9\linewidth]{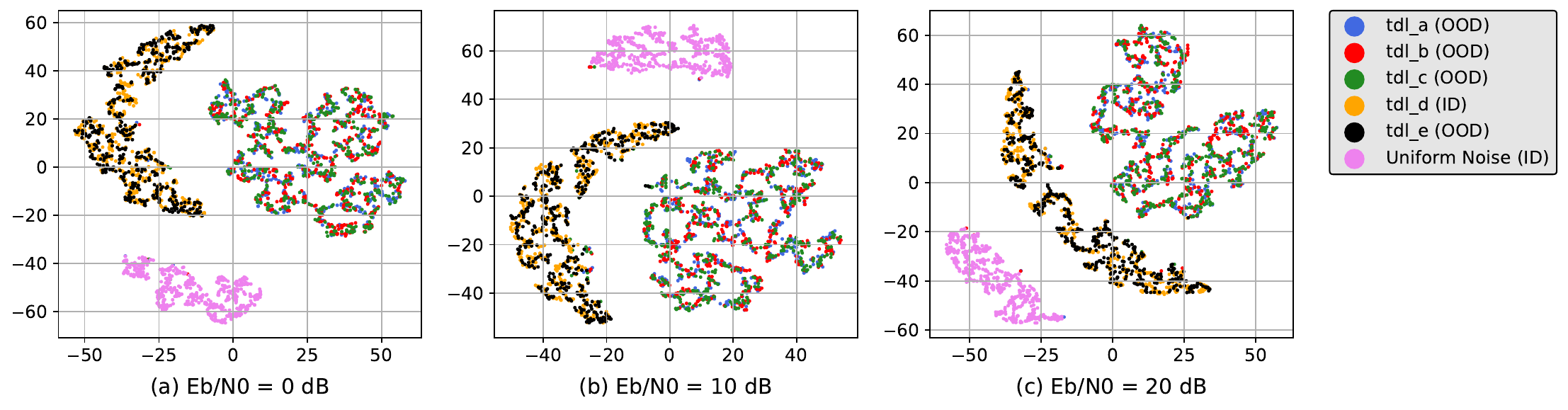}
    \caption{2D projection of feature vectors at the output of Monitor NN for detecting a change in the channel profile. The clusters are shown for different ID and OOD channel profile classes at Eb/N0 levels (a) 0, (b) 10, and (c) 20 dB.}
    \label{fig:clusters-channel}
\end{figure*}
\begin{figure*}[t!!!]
    \centering
    \includegraphics[width=\linewidth]{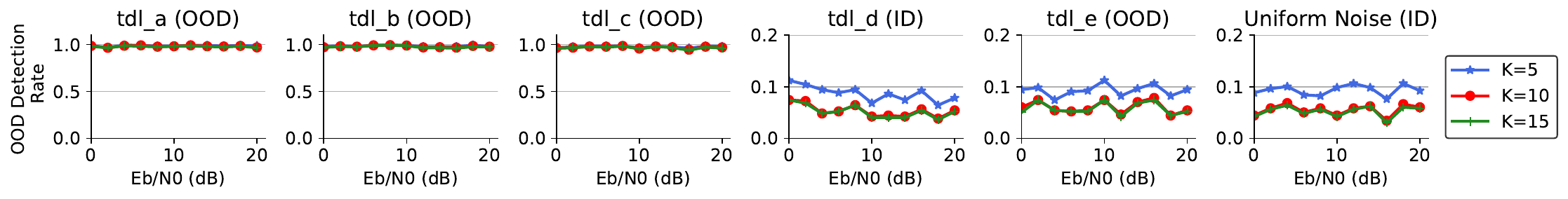}
    \caption{OOD detection rate ($\lambda$ = 0.95) for different test channel profiles with nearest neighbor parameter $K$~= 5, 10, 15.}
    \label{fig:ood-acc-channel}
\end{figure*}
\begin{figure*}[t!!!]
    \centering
    \includegraphics[width=\linewidth]{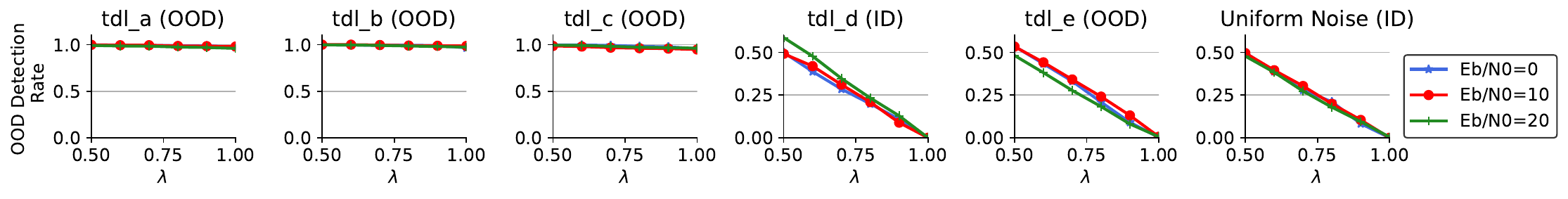}
    \caption{Sensitivity of the proposed OOD detection algorithm (Algorithm~\ref{alg:ood}) to $\lambda$ for different test channel profiles with $K$~=~15 in Eb/N0 levels 0, 10, and 20 dB.}
    \label{fig:channel-lambda-sensitivity}
\end{figure*}

We evaluate the Monitor using dataset configurations guided by DeepRx performance shown in Fig.~\ref{fig:prelim-channel-drop} and summarized in Table~\ref{tab:drop}. As we observe, in DeepRx \emph{performance drop} happens if it is trained on LOS (i.e., tdl\_d) and tested on NLOS (i.e., tdl\_a, tdl\_b, and tdl\_c). Therefore, one desired change detection is the transition between tdl\_d channel profile to either one of the profiles tdl\_a, tdl\_b, or tdl\_c. Based on this, we train the Monitor NN on the training dataset with configurations same as Experiment 2 in Table~\ref{tab:exp_description} that has only tdl\_d channel profile. To be able to construct the triples of \{anchor, positive, negative\} for the triplet loss function, the training set requires to contain more than one class. The second class cannot be based on any of the unseen classes, however, should ideally still have a different distribution from the first class. Therefore, we artificially synthesize a second class derived from the tdl\_d training samples and use it as auxiliary data to train the Monitor NN. To form the second class, we take the pilot matrix for each 5G radio frame for tdl\_d class, and calculate the minimum and maximum values among the all the real and imaginary parts of pilots as $u_{\text{min}}$ and $u_{\text{max}}$, respectively. Then, we create a new matrix with the same dimensions as tdl\_d pilot matrix, and fill it with Uniform noise $\sim U(u_{\text{min}}, u_{\text{max}})$. We train the Monitor NN with triplet loss function to form distinct clusters for tdl\_d and this auxiliary class.

Ideally the fully trained Monitor NN should be able to form distinct clusters for different ID classes and OOD data. The 2D projection of these 256-dimensional clusters is illustrated in Fig.~\ref{fig:clusters-channel} at Eb/N0 levels 0, 10, and 20 dB. We observe that the ID classes tdl\_d (LOS) and Uniform noise form distinct clusters. Regarding the OOD classes, we see that features for all NLOS profiles (i.e., tdl\_a, tdl\_b, and tdl\_c) fall in the same space but form clusters completely distinct from those of the ID classes. We observe that tdl\_e (LOS) features completely overlap with the ID cluster tdl\_d (LOS), since these two channel profiles are very similar, which is also evident in Fig.~\ref{fig:prelim-channel-drop}, where DeepRx model is trained on tdl\_d (LOS), but faces no \emph{performance drop} when it is tested on tdl\_e (LOS). Based on this, the Monitor not being able to distinguish the OOD class tdl\_e from the ID class tdl\_d is not a problem, since DeepRx trained on tdl\_d maintains its higher performance compared to TradRx, when tested on tdl\_e.

The distinct ID and OOD clusters in all low, medium, and high Eb/N0 levels in Fig.~\ref{fig:clusters-channel}(a), (b), and (c), respectively, can be justified by two reasons: First, the pattern of input to the Monitor that is the pseudo random sequence of 5G pilots with QPSK modulation scheme is a simple pattern, and it is not much affected by noise up to Eb/N0~=~0~dB. Second, the Monitor is trained on all Eb/N0 levels in range 0 to 20 dB with steps of 2 dB, and hence, good cluster distinction is observed in all Eb/N0 levels.

In Fig.~\ref{fig:ood-acc-channel}, we observe that the OOD classes tdl\_a, tdl\_b, and tdl\_c achieve 96\%+ OOD detection rate in all Eb/N0 levels for $K$ = 5, 10, 15. The average OOD detection rate for these NLOS channels over all Eb/N0 levels is 97\%+ for all $K$ values. We also observe that the ID classes tdl\_d and Uniform noise achieve a low OOD detection rate (low false positive rate), which is desirable. This low rate reduces from averagely 9.3\% to 5.2\% as we increase $K$ from 5 to 15. We observe that the OOD class tdl\_e does not achieve a high OOD detection rate, however, transitioning from tdl\_d in the training set to tdl\_e during deployment does not cause performance drop for DeepRx as explained above.

Fig.~\ref{fig:channel-lambda-sensitivity} shows OOD detection rate for different channel profile classes vs. $\lambda$ at Eb/N0 levels 0, 10, and 20 dB. We observe that varying $\lambda$ does not impact OOD detection rate for true OOD classes (i.e., tdl\_a, tdl\_b, and tdl\_c), however, for classes that are ID or similar to ID (i.e., tdl\_d, tdl\_e, and Uniform noise) the OOD detection decreases as $\lambda$ increases.

\subsubsection{Detecting a Change in the Transmitter Speed}\label{sec:eval-monitor-speed}
\begin{figure*}[t!!!]
    \centering
    \includegraphics[width=0.9\linewidth]{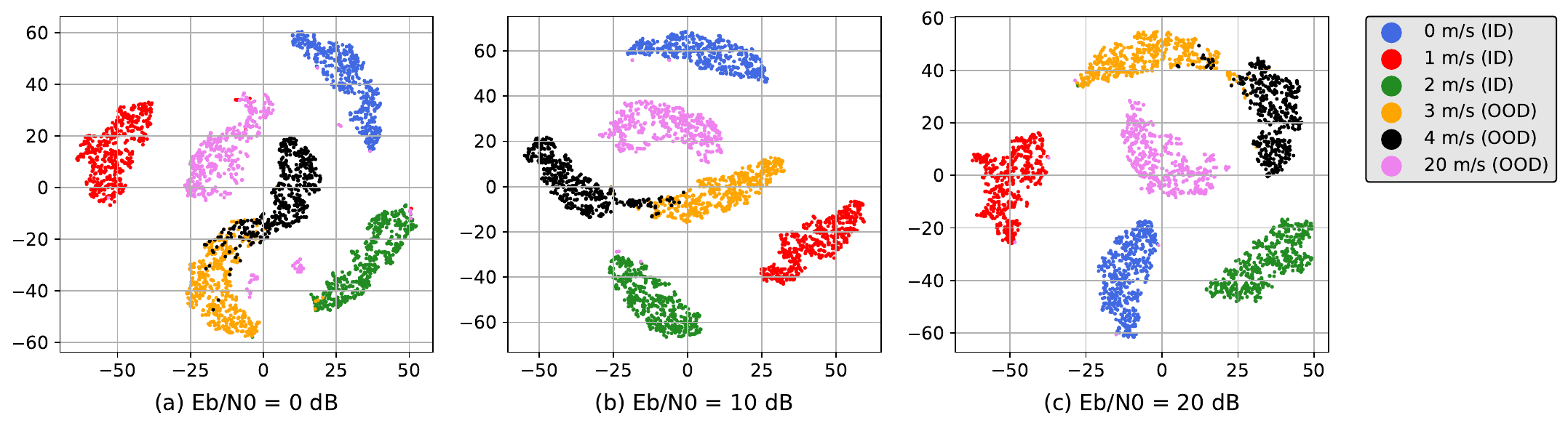}
    \caption{2D projection of feature vectors at the output of Monitor NN for detecting a change in the transmitter speed. The clusters are shown for different ID and OOD transmitter speed classes at Eb/N0 levels (a) 0, (b) 10, and (c) 20 dB.}
    \label{fig:clusters-speed}
\end{figure*}
\begin{figure*}[t!!!]
    \centering
    \includegraphics[width=\linewidth]{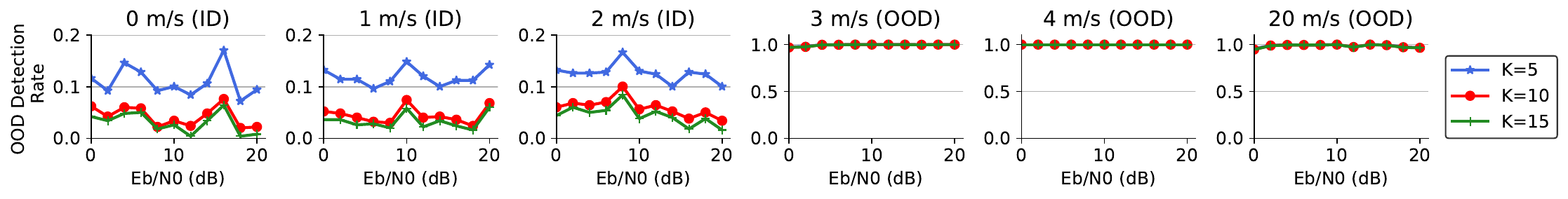}
    \caption{OOD detection rate ($\lambda$ = 0.95) for different test transmitter speeds with nearest neighbor parameter $K$~= 5, 10, 15.}
    \label{fig:ood-acc-speed}
\end{figure*}
\begin{figure*}[t!!!]
    \centering
    \includegraphics[width=\linewidth]{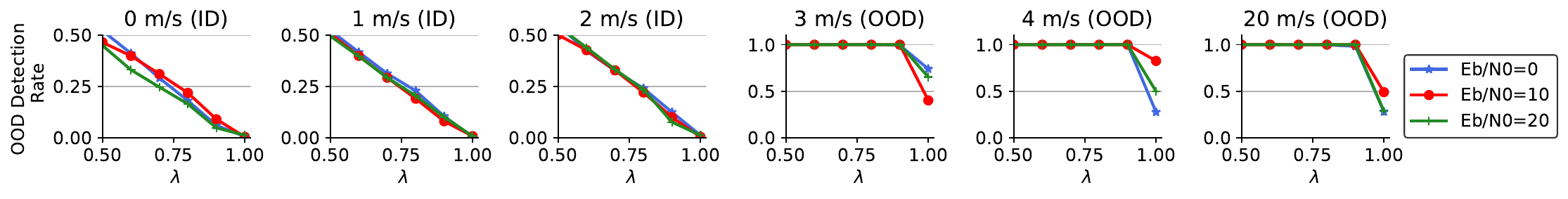}
    \caption{Sensitivity of the proposed OOD detection algorithm (Algorithm~\ref{alg:ood}) to $\lambda$ for different test transmitter speeds with $K$~=~15 in Eb/N0 levels 0, 10, and 20 dB.}
    \label{fig:speed-lambda-sensitivity}
\end{figure*}

Similar to Section~\ref{sec:eval-monitor-channel}, to define ID and OOD transmitter speed classes, we are guided by DeepRx performance shown in Fig.~\ref{fig:prelim-speed-trained-low} and summarized in Table~\ref{tab:drop}. We train the Monitor NN with triplet loss function on the data with the same configurations as the training set of Experiment 4 in Table~\ref{tab:exp_description}, to form distinct clusters for transmitter speeds 0, 1, and 2 m/s.

We test the trained NN on the unseen test set that is a combination of different ID and OOD transmitter speed classes, and visualize 2D projection of their corresponding feature vectors at Eb/N0 levels 0, 10 and 20 dB in Fig.~\ref{fig:clusters-speed}. As observed, all the ID and OOD classes form distinct clusters, even in lower Eb/N0 levels, as explained in Section~\ref{sec:eval-monitor-channel}.

Fig.~\ref{fig:ood-acc-speed} shows OOD detection rate for different ID and OOD transmitter speed classes for nearest neighbor $K$ set as 5, 10, and 15. We observe low OOD detection rate for ID classes 0, 1, and 2 m/s across different Eb/N0 levels, which is desirable as it shows low false positive rate. We see that increasing $K$ from 5 to 15 reduces the average OOD detection rate from 10\% to 3\%, from 11\% to 3\%, and from 12\% to 4\% for ID classes 0, 1, and 2 m/s, respectively. For OOD classes 3, 4, and 20 m/s we observe high OOD detection rate of 98\%+ averaged over all Eb/N0 levels for different $K$ values, which is desirable as it shows high true positive rate.

Fig.~\ref{fig:speed-lambda-sensitivity} shows OOD detection rate vs. $\lambda$ for different transmitter speed classes at Eb/N0 levels 0, 10, and 20 dB. The trend is similar to Fig.~\ref{fig:channel-lambda-sensitivity} in except that OOD detection rate decreases for true OOD classes after a certain $\lambda$ threshold. The implication is when cluster radius increases there is a higher chance that more OOD samples fall inside the cluster and are flagged as ID.

\subsubsection{Detecting a Change in the Delay Spread}\label{sec:eval-monitor-delay}
\begin{figure*}[t!!!]
    \centering
    \includegraphics[width=0.9\linewidth]{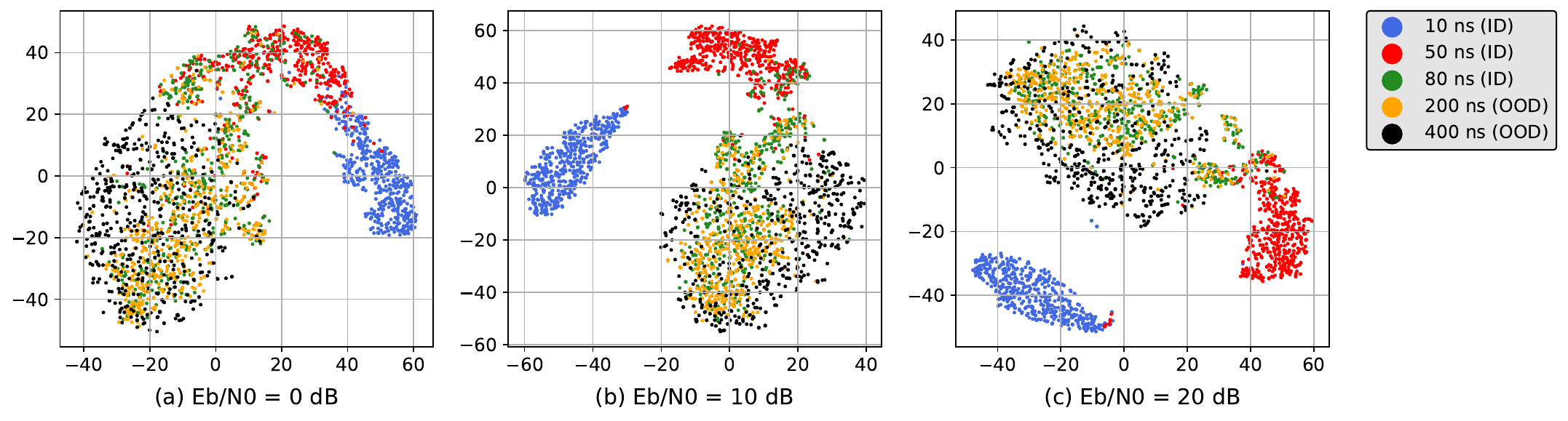}
    \caption{2D projection of feature vectors at the output of Monitor NN for detecting a change in the delay spread. The clusters are shown for different ID and OOD delay spread classes at Eb/N0 levels (a) 0, (b) 10, and (c) 20 dB.}
    \label{fig:clusters-delay}
\end{figure*}
\begin{figure*}[t!!!]
    \centering
    \includegraphics[width=\linewidth]{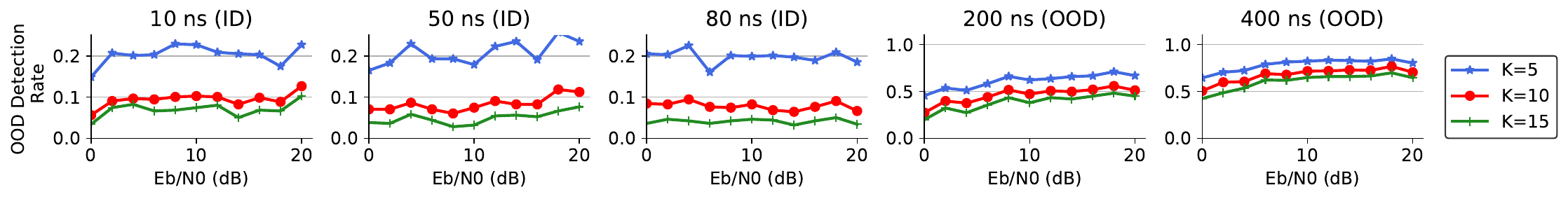}
    \caption{OOD detection rate ($\lambda$ = 0.95) for different test delay spread classes with nearest neighbor parameter $K$~= 5, 10, 15.} 
    \label{fig:ood-acc-delay}
\end{figure*}
\begin{figure*}[t!!!]
    \centering
    \includegraphics[width=\linewidth]{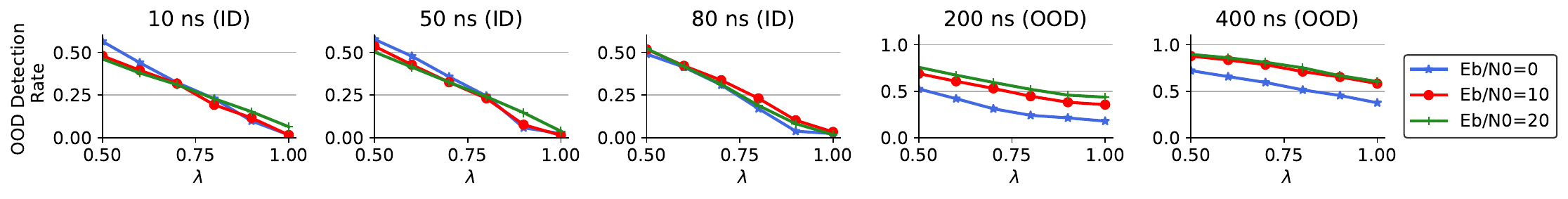}
    \caption{Sensitivity of the proposed OOD detection algorithm (Algorithm~\ref{alg:ood}) to $\lambda$ for different test delay spread classes with $K$~=~15 in Eb/N0 levels 0, 10, and 20 dB.}
    \label{fig:delay-lambda-sensitivity}
\end{figure*}

Similar to Sections~\ref{sec:eval-monitor-channel} and~\ref{sec:eval-monitor-speed} to form ID and OOD classes for Monitor NN, we are guided by DeepRx performance shown in Fig.~\ref{fig:prelim-delay-trained-low} and summarized in Table~\ref{tab:drop}. We train the Monitor NN with triplet loss function on data with the same configurations as the training set of Experiment 6 in Table~\ref{tab:exp_description}, to form distinct clusters for ID delay spread classes 10, 50 and 80 ns.

After training, we test the trained NN on the unseen test set that is a combination of different ID and OOD delay spread classes, and visualize the 2D projection of the generated features for Eb/N0 levels 0, 10, and 20 dB in Fig.~\ref{fig:clusters-delay}.

We observe that with increasing the Eb/N0 level beyond 0 dB, some ID clusters, specially 10 and 50 ns, tend to occupy a more independent space. However, overall the OOD samples are closer to ID clusters 
compared to Figs.~\ref{fig:clusters-channel} and \ref{fig:clusters-speed} and are more difficult to distinguish even in the high Eb/N0 levels. Because of this, we expect to see lower OOD detection rate for OOD classes compared to Sections~\ref{sec:eval-monitor-channel} and \ref{sec:eval-monitor-speed}. 

Fig.~\ref{fig:ood-acc-delay} shows OOD detection rate for different ID and OOD delay spread classes for nearest neighbor $K$ set as 5, 10, and 15. We observe low OOD detection rate for ID classes 10, 50, and 80 ns across different Eb/N0 levels, which is desirable as it shows low false positive rate. We see that increasing $K$ from 5 to 15 reduces the average OOD detection rate from 20\% to 6\%, from 20\% to 4\%, and from 19\% to 4\% for ID classes 10, 50, and 80 ns, respectively. For the OOD classes 200 and 400 ns with $K$~=~5, we observe an increase from 45\% to 66\% and from 19\% to 44\%, respectively, as the Eb/N0 increases from 0 to 20 dB. Furthermore, average OOD detection rate reduces from 61\% to 38\% and from 78\% to 60\%, for classes 200 and 400 ns, respectively, as $K$ increases from 5 to 15. The relatively higher OOD detection rate for the OOD classes despite the partially overlapping clusters for 80, 200, and 400 ns in Fig.~\ref{fig:clusters-delay}, shows that these clusters have some level of separation in the high-dimensional space.

Fig.~\ref{fig:delay-lambda-sensitivity} shows OOD detection rate vs. $\lambda$ for different delay spread classes at Eb/N0 levels 0, 10, and 20 dB. In all ID and OOD classes, we observe OOD detection rate decreases as $\lambda$ (and consequently cluster radius) increases, which is expected for more closely located ID and OOD clusters.

\textbf{Key takeaways.}
\begin{itemize}
    \item \textbf{OOD detection rate relation to 2D visualization.} It is expected that OOD classes that show visual separation from ID clusters in their 2D projection yield higher OOD detection rate, which is observed in Figs.~\ref{fig:ood-acc-channel} and~\ref{fig:ood-acc-speed}.
    \item \textbf{Sensitivity to $K$.} If test feature vectors are closer to ID clusters, as $K$ increases and more neighbors are queried, the chances that at least one neighbor votes for the test feature vector to be ID increases, and hence, the OOD detection rate decreases. This is consistent with tdl\_d, tdl\_e, and Uniform noise plots in Fig.~\ref{fig:ood-acc-channel}, 0, 1, and 2 m/s plots in Fig.~\ref{fig:ood-acc-speed}, and all plots in Fig.~\ref{fig:ood-acc-delay}. However, for test feature vectors that show good separation from ID clusters, we expect that increasing $K$ should not reduce the OOD detection rate, which is consistent with tdl\_a, tdl\_b, and tdl\_c plots in Fig.~\ref{fig:ood-acc-channel} and 3, 4, and 20 m/s plots in Fig.~\ref{fig:ood-acc-speed}.
    \item \textbf{Sensitivity to $\lambda$.} In Figs.~\ref{fig:channel-lambda-sensitivity}, \ref{fig:speed-lambda-sensitivity}, and \ref{fig:delay-lambda-sensitivity}, we observe that high $\lambda$ values ensure low false positive rate for ID classes. For OOD clusters that are well-separated from ID clusters, larger $\lambda$ values close to 1 yield high OOD detection rate. For ODD clusters that are in closer proximity to ID clusters or are partially overlapping with them, slightly lower $\lambda$ values (e.g., 0.90-0.95) can provide an acceptable tradeoff between true and positive rates.    
\end{itemize}

It is worth noting that the experiments conducted in Sections~\ref{sec:eval-monitor-channel},~\ref{sec:eval-monitor-speed}, and~\ref{sec:eval-monitor-delay} are designed to address the corner cases where the Monitor is imposed to as few as possible ID classes during training (i.e., 1-3), and tested on data where only one channel parameter changes (a.k.a., least amount of change that results in near OODs). Increasing the number of ID classes during training or varying multiple wireless channel parameters during test creates an easier OOD detection problem for the Monitor.


\subsection{Performance Comparator}\label{sec:eval-comparator}
To evaluate the Performance Comparator we extract LLRs and calculate bit probabilities for DeepRx and TradRx, for all the test experiments of Channel Profile - Exp. 2, Speed - Exp. 2, and Delay Spread - Exp. 2 in Section~\ref{sec:prelim-drop}. As illustrated in Figs.~\ref{fig:ood-acc-channel}, \ref{fig:ood-acc-speed}, and \ref{fig:ood-acc-delay}, for channel profile, transmitter speed, and delay spread, respectively, any of the inputs with ID or OOD true label might be predicted as OOD, even if it is with a low probability in the case of inputs with ID true labels. Therefore, the proposed Performance Comparator must be able to correctly trigger retraining for not only for OOD classes but also the ID classes. To evaluate this, we define an accuracy metric for the Performance Comparator based on the output from Algorithm~\ref{alg:comparator}. 
We evaluate the Performance Comparator on a \emph{per-frame} basis, which means we collect the LLRs from DeepRx and TradRx for one 5G radio frame with 6 PBRs per subframe that is equivalent to 36k softbits from each receiver, and feed them to Algorithm~\ref{alg:comparator}. The 36k that is the number of softbits in each 5G radio frame is achieved through the following calculations:
\begin{align*}
&10 \text{ [subframes]} \times 4 \text{ [16QAM modulation]} \times \\
&(72\times14-36\times3) \text{[data minus pilots]}\\ &= 36000 \text{ [softbits]}
\end{align*}       

The Performance Comparator determines whether or not DeepRx needs retraining once for every 5G radio frame. In Algorithm~\ref{alg:comparator}, we consider each prediction as a correct decision if the Performance Comparator flags retraining as ``needed'' and in fact $\text{BER}_\text{DeepRx} > \text{BER}_\text{TradRx}$ for the corresponding frame, or if it flags retraining as ``not needed'' and in fact $\text{BER}_\text{DeepRx} \leq \text{BER}_\text{TradRx}$ for that corresponding frame. For each test set, we calculate Performance Comparator accuracy as the number of correct decisions divided by the total number of decisions. We show Performance Comparator accuracy for different ID and OOD test sets of different channel profiles, transmitter speeds, and delay spreads in Sections~\ref{sec:eval-comparator-channel}, \ref{sec:eval-comparator-speed}, and \ref{sec:eval-comparator-delay}, respectively.

\subsubsection{Performance Comparator Accuracy in Different Channel Profiles}\label{sec:eval-comparator-channel}
\begin{figure}[t!!!]
\centering
\begin{flushleft}
\begin{tikzpicture}
        \begin{axis}[
            axis lines = left,
            xlabel = Eb/N0 (dB),
            ylabel = Comparator Accuracy,
            ymin=0,
            ymax=1.1,
            xmin=0,
            xmax=20,
            height=4.5cm,
            width=7.5cm,
            xtick={0,5,10,15,20},
            legend style={at={(0,0)},
            fill=gray!10,
            fill opacity=1.0,
            text opacity=1,
            draw=none,
            font=\small,
            legend cell align=left,
                anchor=south west,legend columns=2},
            ymajorgrids=true,
            ytick = {0,0.25,0.5,0.75,1.0},
            grid style={line width=1pt,draw=gray!50},
        ]
        \addplot [very thick,blue,mark=square*]
        coordinates{(0,1)(2,1)(4,1)(6,0.998)(8,0.934)(10,0.748)(12,0.458)(14,0.314)(16,0.454)(18,0.686)(20,0.91)};
        \addlegendentry{tdl\_a}
        \addplot [very thick,red,mark=x,mark size=4pt]
        coordinates{(0,1)(2,1)(4,1)(6,0.992)(8,0.948)(10,0.796)(12,0.518)(14,0.33)(16,0.38)(18,0.652)(20,0.9)};
        \addlegendentry{tdl\_b}
        \addplot [very thick,ForestGreen,mark=*]
        coordinates{(0,1)(2,1)(4,1)(6,0.992)(8,0.95)(10,0.852)(12,0.618)(14,0.356)(16,0.412)(18,0.602)(20,0.88)};
        \addlegendentry{tdl\_c}
        \addplot [very thick,cyan,mark=diamond*,mark size=3]
        coordinates{(0,1)(2,1)(4,1)(6,1)(8,1)(10,1)(12,1)(14,0.998)(16,0.996)(18,0.986)(20,0.958)};
        \addlegendentry{tdl\_d}
        \addplot [very thick,magenta,mark=|,mark size=4pt]
        coordinates{(0,1)(2,1)(4,1)(6,1)(8,1)(10,1)(12,1)(14,1)(16,0.992)(18,0.988)(20,0.962)};
        \addlegendentry{tdl\_e}
        \end{axis}
        \begin{axis}[
            at={(6.5cm,0)},   
            axis lines = left,
            axis y line*=left,
            axis x line=none,
            anchor=south west,
            width=2.2cm,
            height=4.5cm,
            xmin=0, xmax=1,
            ymin=0, ymax=1.1,
            title=Avg.,
            title style = {yshift=-3.7cm},
            xtick=\empty,
            ymajorgrids=true,
            ytick = {0,0.25,0.5,0.75,1.0},
            yticklabels={},
            grid style={line width=1pt,draw=gray!50},
        ]
        \addplot [very thick, blue,mark=square*] coordinates {(0.5,0.773)};
        \addplot [very thick, red, mark=x,mark size=4pt] coordinates {(0.5,0.774)};
        \addplot [very thick, ForestGreen, mark=*] coordinates {(0.5,0.78)};
        \addplot [very thick, cyan, mark=diamond*,mark size=3] coordinates {(0.5,0.99)};
        \addplot [very thick, magenta, mark=|,mark size=4pt] coordinates {(0.5,0.99)};
        \end{axis}
    \end{tikzpicture}
    \end{flushleft}
    \caption{Comparator accuracy when 5G radio frames with different channel profiles are passed through the TradRx and DeepRx trained on channel profile tdl\_d. The plot on the right shows accuracy averaged over Eb/N0 levels.}
    \label{fig:eval-comparator-channel}
\end{figure}
In Fig.~\ref{fig:eval-comparator-channel}, where DeepRx is trained on tdl\_d channel profile, we observe Performance Comparator accuracies of 77\%, 77\%, 78\%, 99\%, and 99\%, averaged over all Eb/N0 levels, for test channel profiles tdl\_a, tdl\_b, tdl\_c, tdl\_d, and tdl\_e, respectively. This can be averaged to 86\% accuracy for all the test channel profiles in all Eb/N0 levels. Lowest accuracy in range 31-35\% can be seen for the NLOS channel profiles tdl\_a, tdl\_b, and tdl\_c, in Eb/N0=14 dB. Comparing this with Fig.~\ref{fig:prelim-channel-drop} shows the low accuracy happens close to the Eb/N0 level where the two BER graphs of DeepRx and TradRx cross (i.e., 12 dB). This means the Performance Comparator makes incorrect decisions mostly when $\text{BER}_\text{DeepRx} \approx \text{BER}_\text{TradRx}$. At Eb/N0=14 dB, $\text{BER}_\text{DeepRx}$ is only 2.6e-3  higher than $\text{BER}_\text{TradRx}$. Therefore, an incorrect decision of ``retraining not required'' hurts the system BER a negligible amount of 2.6e-3 higher BER, for $\sim$70\% of the frames.

\subsubsection{Performance Comparator Accuracy in Different Transmitter Speeds}\label{sec:eval-comparator-speed}
\begin{figure}[t!!!]
\centering
\begin{tikzpicture}
        \begin{axis}[
            axis lines = left,
            xlabel = Eb/N0 (dB),
            ylabel = Comparator Accuracy,
            ymin=0,
            ymax=1.1,
            xmin=0,
            xmax=20,
            height=4.5cm,
            width=7.5cm,
            xtick={0,5,10,15,20},
            xticklabels = {0,5,10,15,20},
            legend style={at={(1,0)},
            fill=gray!10,
            fill opacity=1.0,
            text opacity=1,
            draw=none,
            font=\small,
            legend cell align=left,
                anchor=south east,legend columns=2},
            ymajorgrids=true,
            ytick = {0,0.25,0.5,0.75,1.0},
            yticklabels = {0,0.25,0.5,0.75,1.0},
            grid style={line width=1pt,draw=gray!50},
        ]
        \addplot [very thick,blue,mark=square*]
        coordinates{
        (0,1)(2,1)(4,1)(6,1)(8,1)(10,1)(12,1)(14,0.998)(16,0.944)(18,0.902)(20,0.952)
        };
        \addlegendentry{0 m/s}
        \addplot [very thick,red,mark=x,mark size=4pt]
        coordinates{
        (0,1)(2,1)(4,1)(6,1)(8,1)(10,1)(12,1)(14,0.998)(16,0.954)(18,0.916)(20,0.948)
        };
        \addlegendentry{1 m/s}
        \addplot [very thick,ForestGreen,mark=*]
        coordinates{
        (0,1)(2,1)(4,1)(6,1)(8,1)(10,1)(12,1)(14,0.996)(16,0.94)(18,0.896)(20,0.93)
        };
        \addlegendentry{2 m/s}
        \addplot [very thick,cyan,mark=diamond*,mark size=3]
        coordinates{
        (0,1)(2,1)(4,1)(6,1)(8,1)(10,1)(12,1)(14,0.994)(16,0.946)(18,0.926)(20,0.95)
        };
        \addlegendentry{3 m/s}
        \addplot [very thick,magenta,mark=|,mark size=4pt]
        coordinates{
        (0,1)(2,1)(4,1)(6,1)(8,1)(10,1)(12,1)(14,0.994)(16,0.88)(18,0.848)(20,0.934)
        };
        \addlegendentry{4 m/s}
        \addplot [very thick,gray,mark=+,mark size=4pt]
        coordinates{
        (0,1)(2,0.988)(4,0.11)(6,0)(8,0.148)(10,0.998)(12,1)(14,1)(16,1)(18,1)(20,1)
        };
        \addlegendentry{20 m/s}
        \end{axis}
        \begin{axis}[
            at={(6.5cm,0)},   
            axis lines = left,
            axis y line*=left,
            axis x line=none,
            anchor=south west,
            width=2.2cm,
            height=4.5cm,
            xmin=0, xmax=1,
            ymin=0, ymax=1.1,
            title=Avg.,
            title style = {yshift=-3.7cm},
            xtick=\empty,
            ymajorgrids=true,
            ytick = {0,0.25,0.5,0.75,1.0},
            yticklabels={},
            grid style={line width=1pt,draw=gray!50},
        ]
        \addplot [very thick, blue,mark=square*] coordinates {(0.5,0.98)};
        \addplot [very thick, red, mark=x,mark size=4pt] coordinates {(0.5,0.98)};
        \addplot [very thick, ForestGreen, mark=*] coordinates {(0.5,0.97)};
        \addplot [very thick, cyan, mark=diamond*,mark size=3] coordinates {(0.5,0.98)};
        \addplot [very thick, magenta, mark=|,mark size=4pt] coordinates {(0.5,0.96)};
        \addplot [very thick, gray, mark=+,mark size=4pt] coordinates {(0.5,0.75)};
        \end{axis}
    \end{tikzpicture}
    \caption{Comparator accuracy when 5G radio frames with different transmitter speeds are passed through the TradRx and DeepRx trained on transmitter speeds 0, 1, and 2 m/s. The plot on the right shows accuracy averaged over Eb/N0 levels.}
    \label{fig:eval-comparator-speed}
\end{figure}
In Fig.~\ref{fig:eval-comparator-speed}, where DeepRx is trained on speeds 0, 1, and 2 m/s, we observe comparator accuracies of 98\%, 98\%, 97\%, 98\%, 96\%, and 75\%, averaged over all Eb/N0 levels, for test speeds 0, 1, 2, 3, 4, and 20 m/s, respectively. This can be averaged to 93.3\% accuracy for all the test speeds in all Eb/N0 levels. Lowest accuracy of 0\% can be seen for the highest speed of 20 m/s in Eb/N0=6 dB. Comparing this with Fig.~\ref{fig:prelim-speed-trained-low} shows the low accuracy happens close to the Eb/N0 level that the two BER graphs of DeepRx and TradRx cross (i.e., 4 dB). Similar to Section~\ref{sec:eval-comparator-channel}, the comparator makes incorrect decisions mostly when $\text{BER}_\text{DeepRx} \approx \text{BER}_\text{TradRx}$. At Eb/N0=6 dB for speed 20 m/s, $\text{BER}_\text{DeepRx}$ is only 1.4e-2 higher than $\text{BER}_\text{TradRx}$. Therefore, an incorrect decision of ``retraining not required'' hurts the system BER by only 1.4e-2 higher BER.

\subsubsection{Performance Comparator Accuracy in Different Delay Spreads}\label{sec:eval-comparator-delay}
\begin{figure}[t!!!]
\centering
\begin{tikzpicture}
        \begin{axis}[
            axis lines = left,
            xlabel = Eb/N0 (dB),
            ylabel = Comparator Accuracy,
            ymin=0,
            ymax=1.1,
            xmin=0,
            xmax=20,
            height=4.5cm,
            width=7.5cm,
            xtick={0,5,10,15,20},
            xticklabels = {0,5,10,15,20},
            legend style={at={(1,0)},
            fill=gray!10,
            fill opacity=1.0,
            text opacity=1,
            draw=none,
            font=\small,
            legend cell align=left,
                anchor=south east,legend columns=2},
            ymajorgrids=true,
            ytick = {0,0.25,0.5,0.75,1.0},
            yticklabels = {0,0.25,0.5,0.75,1.0},
            grid style={line width=1pt,draw=gray!50},
        ]
        \addplot [very thick,blue,mark=square*]
        coordinates{
        (0,1.0)(2,1.0)(4,1.0)(6,1.0)(8,1.0)(10,1.0)(12,1.0)(14,1.0)(16,0.986)(18,0.972)(20,0.97)
        };
        \addlegendentry{10 ns}
        \addplot [very thick,red,mark=x,mark size=4pt]
        coordinates{
       (0,1.0)(2,1.0)(4,1.0)(6,1.0)(8,1.0)(10,1.0)(12,1.0)(14,0.998)(16,0.99)(18,0.98)(20,0.974)
        };
        \addlegendentry{50 ns}
        \addplot [very thick,ForestGreen,mark=*]
        coordinates{
        (0,1)(2,1)(4,1)(6,1)(8,1)(10,1)(12,1)(14,0.996)(16,0.986)(18,0.982)(20,0.986)
        };
        \addlegendentry{80 ns}
        \addplot [very thick,cyan,mark=diamond*,mark size=3]
        coordinates{
        (0,1)(2,1)(4,1)(6,1)(8,1)(10,0.992)(12,0.968)(14,0.922)(16,0.902)(18,0.85)(20,0.86)
        };
        \addlegendentry{200 ns}
        \addplot [very thick,magenta,mark=|,mark size=4pt]
        coordinates{
        (0,0.974)(2,0.916)(4,0.854)(6,0.728)(8,0.65)(10,0.62)(12,0.732)(14,0.81)(16,0.9)(18,0.926)(20,0.98)
        };
        \addlegendentry{400 ns}
        \end{axis}
        \begin{axis}[
            at={(6.5cm,0)},   
            axis lines = left,
            axis y line*=left,
            axis x line=none,
            anchor=south west,
            width=2.2cm,
            height=4.5cm,
            xmin=0, xmax=1,
            ymin=0, ymax=1.1,
            title=Avg.,
            title style = {yshift=-3.7cm},
            xtick=\empty,
            ymajorgrids=true,
            ytick = {0,0.25,0.5,0.75,1.0},
            yticklabels={},
            grid style={line width=1pt,draw=gray!50},
        ]
        \addplot [very thick, blue,mark=square*] coordinates {(0.5,0.99)};
        \addplot [very thick, red, mark=x,mark size=4pt] coordinates {(0.5,0.99)};
        \addplot [very thick, ForestGreen, mark=*] coordinates {(0.5,0.99)};
        \addplot [very thick, cyan, mark=diamond*,mark size=3] coordinates {(0.5,0.95)};
        \addplot [very thick, magenta, mark=|,mark size=4pt] coordinates {(0.5,0.82)};
        \end{axis}
    \end{tikzpicture}
    \caption{Comparator accuracy when 5G radio frames with different delay spreads are passed through the TradRx and DeepRx trained on delay spreads 10, 50, and 80 ns. The plot on the right shows accuracy averaged over Eb/N0 levels.}
    \label{fig:eval-comparator-delay}
\end{figure}
In Fig.~\ref{fig:eval-comparator-delay}, where DeepRx is trained on delay spreads 10, 50, and 80 ns, we observe comparator accuracies of 99\%, 99\%, 99\%, 95\%, and 82\%, averaged over all Eb/N0 levels, for test delay spreads 10, 50, 80, 200, and 400 ns, respectively. This can be averaged to 94.8\% accuracy for all test delay spreads in all Eb/N0 levels. Lowest accuracy of 62\% can be seen for the highest delay spread of 400 ns in Eb/N0=10 dB. Comparing this with Fig.~\ref{fig:prelim-delay-trained-low} shows the low accuracy happens close to the Eb/N0 level where the two BER graphs of DeepRx and TradRx cross (i.e., 6 dB). Similar to Sections~\ref{sec:eval-comparator-channel} and \ref{sec:eval-comparator-speed}, the comparator makes incorrect decisions mostly when $\text{BER}_\text{DeepRx} \approx \text{BER}_\text{TradRx}$. At Eb/N0=10 dB for delay spread 400 ns, $\text{BER}_\text{DeepRx}$ is only 1.0e-2 higher than $\text{BER}_\text{TradRx}$. Therefore, an incorrect decision of ``retraining not required'' hurts the system BER by only 1.0e-2 higher BER for only $\sim$38\% of the 5G radio frames.

\section{Discussion}\label{sec:discussion}

In this section, we study the efficacy of VERITAS in terms of the computational overhead that it imposes to the AI-native receiver system, and we compare it against naive periodic retraining of the AI-native receiver.

The only component in VERITAS that is always active and continuously runs is the Monitor. The Monitor NN and the OOD detection algorithm have runtimes of 537 $\mu$s on Nvidia RTX 6000 GPU and 592 $\mu$s on CPU, respectively, per input. The total runtime for the Monitor entity adds up to averagely $\sim$1.13 ms that is shorter than the duration of three radio frames (i.e., 30 ms) that construct the Monitor input, by a large margin. This shows that the Monitor can operate on streaming 5G frames, even if implemented in software. To avoid being limited to hardware and implementation-dependent runtime metrics, we focus on algorithmic level complexity for comparison and report the number of floating point operations (FLOPs) in the rest of this section.

We use Python package \texttt{ptflops} to count inference FLOPs for the Monitor NN as a \texttt{PyTorch} model. We also analyze different steps and equations that construct the OOD detection algorithm, the Performance Comparator, and the TradRx, and calculate FLOPs for each component in the VERITAS system as shown in Table~\ref{tab:flops-a}.

\begin{table}[t!!!]
    \begin{minipage}{0.4\linewidth}
    \resizebox{\linewidth}{!}{
    \begin{tabular}{||l|c||}
        \hline
        \xrowht[()]{7pt}
        VERITAS &  Inference  \\  
        Components &  GFLOPs \\  
        \hline \hline
        \xrowht[()]{7pt}
        Monitor NN & 1.181 \\  
        \hline
        \xrowht[()]{7pt}
        OOD Detection & \multirow{2}{*}{0.012} \\ 
        Algorithm (K=15) & \\
        \hline
        \xrowht[()]{7pt}
        Monitor Total & 1.193\\
        \hline
        \xrowht[()]{7pt}
        Comparator & 0.005 \\ 
        \hline
        \xrowht[()]{7pt}
        TradRx & 0.004 \\ 
        \hline
        \xrowht[()]{7pt}
        \textbf{VERITAS Total} & \textbf{1.202} \\
        \hline
    \end{tabular}
    }
    \subcaption{}\label{tab:flops-a}
    \end{minipage}\hfill
    \begin{minipage}{0.53\linewidth}
    \resizebox{\linewidth}{!}{
    \begin{tabular}{||l|c||}
        \hline
        \xrowht[()]{7pt}
        DeepRx &  GFLOPs \\  
        \hline \hline
        \xrowht[()]{7pt}
        Inference on & \multirow{2}{*}{1.330} \\
        one radio frame & \\
        \hline
        \xrowht[()]{7pt}
        Optimizer & 0.013 \\ 
        \hline
        \xrowht[()]{7pt}
        Training on & \multirow{2}{*}{4.003} \\
        one radio frame & \\
        \hline
        \xrowht[()]{7pt}
        Training on 300 radio & \multirow{2}{*}{6004.5} \\
        frames for 5 epochs & \\
        \hline
    \end{tabular}
    }
    \subcaption{}\label{tab:flops-b}
    \end{minipage}
    \caption{(a) The number of FLOPs for different components in VERITAS, computed for one input with size (2, 90, 30) corresponding to three 5G frames as explained in Fig.~\ref{fig:monitor-arch}. (b) DeepRx FLOPs in different modes.}\label{tab:flops}
\end{table}

Furthermore, we use Python package \texttt{keras\_flops} to calculate FLOPs for training the AI-native receiver (i.e., DeepRx) using LAMB optimizer~\cite{lamb}. We achieve FLOP count of 4.003 G for training DeepRx on one 5G radio frame. We consider the case where a change in the channel conditions occurs and DeepRx needs to be retrained. As we show in Section~\ref{sec:preliminaries} training DeepRx from scratch requires $\sim$5000 radio frames per Eb/N0, channel profile, transmitter speed, and delay spread, and continues for 20 epochs. We assume a minimum of $\sim$300 radio frames and a minimum of 5 epochs required to retrain DeepRx in the field and fine-tune it on the new channel conditions. In this case, the total FLOPs for retraining DeepRx is estimated as $300\times4.003\times10^9\times5 = 6004.5$ GFLOPs, as shown in Table~\ref{tab:flops-b}.




VERITAS aims to replace naive periodic retraining of the AI-native receiver to avoid unnecessary retraining, as discussed in Sections~\ref{sec:intro} and~\ref{sec:related-maintenace}. In periodic retraining, the frequency of retraining is configurable by the designer, and is upper bounded by the number of radio frames needed to retrain the AI-native receiver. In our AI-native example (i.e., DeepRx) in extreme cases and hypothetical setting, retraining can happen as frequent as every 300 radio frames (i.e., every 3 seconds) as discussed above.


According to Table~\ref{tab:flops}, the FLOPs count for retraining DeepRx matches the FLOPs of the Monitor running (6004.5/1.193$\approx$) 5000 times. This is equivalent to the Monitor processing (5000$\times$3$\approx$) 15000 radio frames that have a total time duration of 150 seconds. VERITAS does not provide lower computational complexity compared to periodic retraining, if DeepRx is scheduled to retrain less frequently than every $\sim$150 seconds. However, VERITAS provides lower computational complexity, if DeepRx periodic retraining happens more frequently than every $\sim$150 seconds. For example, assuming that periodic retraining of DeepRx is scheduled for every 3 seconds (a.k.a., the most frequent retraining possible), the Monitor has to run 100 times to process 300 frames. In this case, the total FLOPs for running the Monitor add up to (100$\times$1.193=) 119.3 GFLOPs, which is only (119.3/6004.5$\approx$) 2\% the computational complexity of retraining DeepRx.

It should be emphasized that retraining of AI-native receiver requires not only dedicated compute resources, but also signals collected under the current channel with known transmit bit labels, whose collection is a significant communication overhead. VERITAS helps reduce both training computation and communication overhead while preventing AI-native receiver performance degradation caused by environment variations.

\section{Conclusion}\label{sec:conclusion}
In this paper, we proposed VERITAS as a framework for verifying the performance of AI-native receivers. VERITAS consists of a Monitor, a Performance Comparator, and a traditional receiver as the reference point. The Monitor that is an OOD detector NN constantly observes the wireless channel and detects changes in different parameters: channel profile (i.e., LOS or NLOS environment), transmitter speed, and delay spread. The proposed Monitor shows 99\%, 97\%, and 69\% true OOD detection rate for channel profile, transmitter speed, and delay spread, respectively. As soon as a change in the wireless channel is detected, the Monitor activates a TradRx to be used as a reference receiver that runs in parallel to the NN-based receiver. The Performance Comparator compares the bit probabilities yielding from the same data inputs passing through DeepRx and TradRx and identifies the receiver with higher BER, to determine whether or not a retraining process needs to be started. The proposed Performance Comparator correctly triggers retraining with an average accuracy of 86\%, 93.3\%, and 94.8\% for all channel profile, transmitter speed, and delay spread test sets, averaged over all Eb/N0 levels. VERITAS can assist in verifying and maintaining the BER performance of AI-native receivers in real-world deployments, such as the real-time AI-native receiver prototype integrated into NI's USRP-based research platform described in~\cite{ni-white-deeprx}.

\section*{Acknowledgment}
This work is supported by U.S. National Science Foundation under grants CNS-2526493 and CNS-2112471.

\bibliographystyle{ieeetr} 
\bibliography{ref} 

\begin{IEEEbiography}
[{\includegraphics[width=1in,height=1.5in,clip,keepaspectratio]{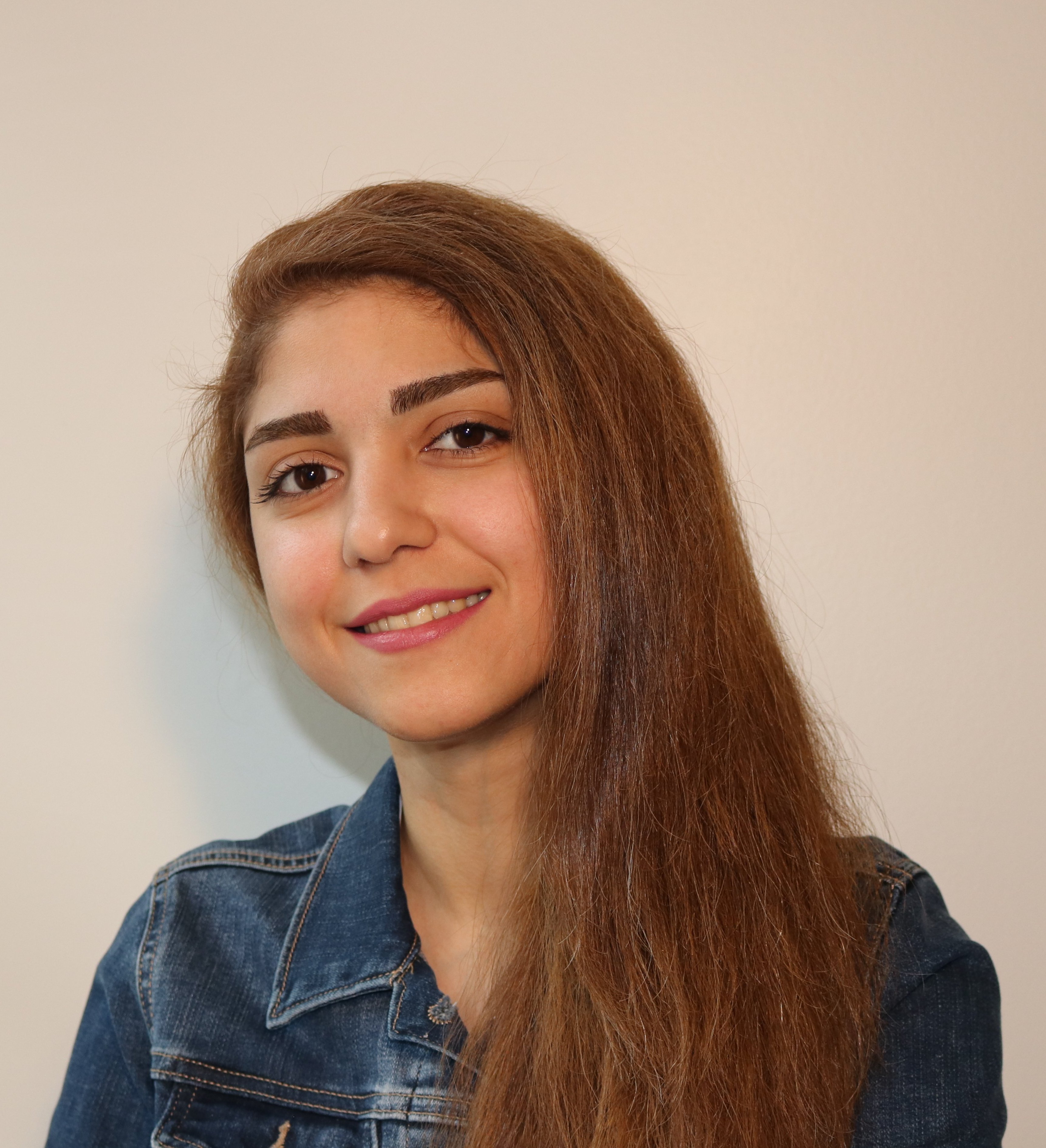}}] 
{Nasim Soltani} (Student Member, IEEE) is a PhD candidate at the Wireless Networking and Communication Group (WNCG) under electrical and computer engineering department at The University of Texas at Austin. Her research interest is broadly applied AI/ML for wireless communications. She has used deep learning algorithms in her research in different domains including signal classification, RF fingerprinting, and AI-native receivers, for different wireless technologies including WiFi and cellular.
\end{IEEEbiography}
\vskip -2\baselineskip plus -1fil

\begin{IEEEbiography}
[{\includegraphics[width=1in,height=1.5in,clip,keepaspectratio]{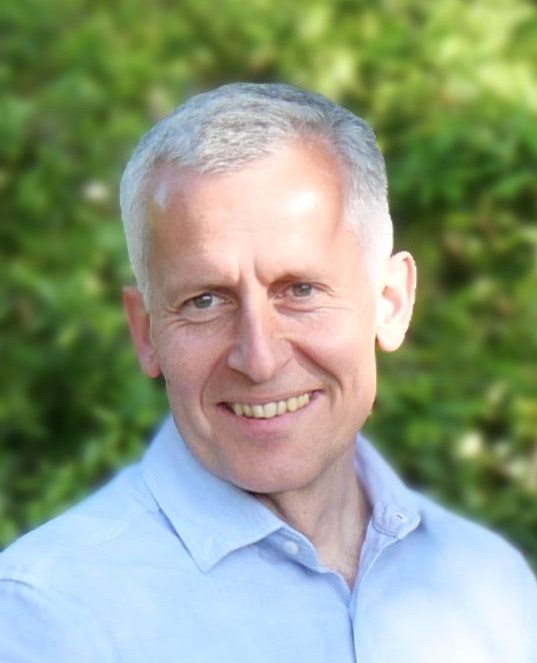}}] 
{Michael L{\"o}hning} received the M.Sc. and Ph.D. degrees in Electrical Engineering from Dresden University of Technology, Germany, in 1999 and 2006, respectively. He is Chief Engineer at NI, based in Dresden, Germany. His research interests focus on emerging technologies for future wireless communication systems and RF test solutions, including the application and validation of AI/ML approaches. 
\end{IEEEbiography}
\vskip -2\baselineskip plus -1fil

\begin{IEEEbiography}
[{\includegraphics[width=1in,height=1.25in,clip,keepaspectratio]{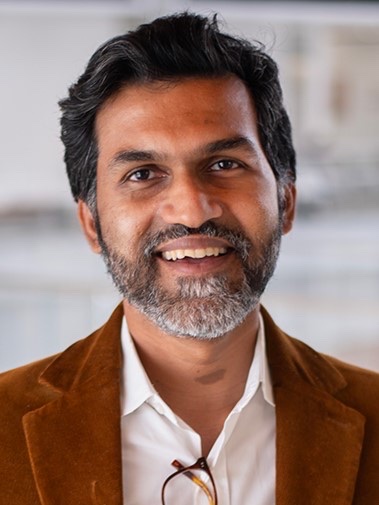}}]{Kaushik Chowdhury} (Fellow, IEEE) received the PhD degree from the Georgia Institute of Technology, in 2009. He is a Chandra Family Endowed distinguished professor in electrical and computer engineering with The University of Texas at Austin. His current research interests involve systems aspects of machine learning for agile spectrum sensing/access, unmanned autonomous systems, programmable and open cellular networks, and largescale experimental deployment of emerging wireless technologies.
\end{IEEEbiography}

\end{document}